\documentclass[prx, twocolumn, superscriptaddress]{revtex4-2}
\usepackage{bm, amsmath, amsfonts, amssymb, mathtools, braket}
\usepackage{multirow}
\usepackage{graphicx}
\usepackage{float, color, xcolor}

\usepackage{bbm}
\usepackage{tabularx}

\usepackage[
pagebackref=false,
colorlinks=true,
linkcolor=blue,
urlcolor=blue,
filecolor=black,
citecolor=red,
pdfstartview=FitV,
pdftitle={},
pdfauthor={},
pdfsubject={},
pdfkeywords={},
pdfpagemode=None,
bookmarksopen=true
]{hyperref}

\newcommand{\ii}{\text{i}}
\newcommand{\U}{U}

\begin{document}

\title{Singular-Value Statistics of \\ 
Non-Hermitian Random Matrices and Open Quantum Systems}

\author{Kohei Kawabata}
\email{kawabata@issp.u-tokyo.ac.jp}
\affiliation{Department of Physics, Princeton University, Princeton, New Jersey 08544, USA}
\affiliation{Institute for Solid State Physics, University of Tokyo, Kashiwa, Chiba 277-8581, Japan}

\author{Zhenyu Xiao}
\affiliation{International Center for Quantum Materials, Peking University, Beijing 100871, China}

\author{Tomi Ohtsuki}
\affiliation{Physics Division, Sophia University, Chiyoda-ku, Tokyo 102-8554, Japan}

\author{Ryuichi Shindou}
\affiliation{International Center for Quantum Materials, Peking University, Beijing 100871, China}

\date{\today}

\begin{abstract}
The spectral statistics of non-Hermitian random matrices are of importance as a diagnostic tool for chaotic behavior in open quantum systems.
Here, we investigate the statistical properties of singular values in non-Hermitian random matrices as an effective measure of quantifying dissipative quantum chaos.
By means of Hermitization, we reveal the unique characteristics of the singular-value statistics that distinguish them from the complex-eigenvalue statistics, and establish the comprehensive classification of the singular-value statistics for all the 38-fold symmetry classes of non-Hermitian random matrices.
We also analytically derive the singular-value statistics of small random matrices, 
which well describe those of large random matrices in the similar spirit to the Wigner surmise.
Furthermore, we demonstrate that singular values of open quantum many-body systems follow the random-matrix statistics, thereby identifying chaos and nonintegrability in open quantum systems.
Our work elucidates that the singular-value statistics serve as a clear indicator of symmetry and lay a foundation for statistical physics of open quantum systems.
\end{abstract}

\maketitle

\section{Introduction}

The spectral statistics serve as a defining feature of chaotic behavior in quantum systems and lack thereof, and play a fundamental role in the principles of statistical mechanics~\cite{Haake-textbook, Huse-review, Rigol-review}.
According to a number of numerical and analytical investigations, it is widely accepted that the spectrum of a nonintegrable quantum system follows the random-matrix statistics~\cite{BGS-84}, while the spectrum of an integrable quantum system follows the Poisson statistics~\cite{Berry-Tabor-77}.
Here, the spectral statistics of Hermitian random matrices are universally determined by the 10-fold symmetry classification~\cite{AZ-97}.
These 10 symmetry classes encompass the 3 standard (Wigner-Dyson) classes by time-reversal symmetry~\cite{Wigner-51, *Wigner-58, Dyson-62}, 3 chiral classes by chiral symmetry~\cite{Gade-91, *Gade-93, Verbaarschot-94, *Verbaarschot-00-review}, and 4 Bogoliubov-de Gennes classes by particle-hole symmetry~\cite{AZ-97}.
Time-reversal symmetry influences the correlations in the bulk of the spectra, quantified by the Dyson index $\beta = 1, 2, 4$.
On the other hand, the impact of chiral or particle-hole symmetry arises only around the symmetric point (i.e., zero eigenvalue) and is classified by the other random-matrix index $\alpha = 0, 1, 2, 3$.
This 10-fold symmetry classification is also fundamental in understanding mesoscopic transport phenomena including the Anderson transitions~\cite{Beenakker-review-97, *Beenakker-review-15, Evers-review}, as well as topological insulators and superconductors~\cite{Schnyder-08, *Ryu-10, Kitaev-09, HK-review, QZ-review, CTSR-review}.

Recently, there has been a remarkable surge of interest in the physics of open quantum systems. 
As coupling with the surrounding environment is an inevitable aspect of realistic physical systems, an understanding of open quantum systems is important for quantum simulation and technology.
In contrast to closed quantum systems, open quantum systems are no longer described by Hermitian Hamiltonians.
In general, the dynamics of open quantum systems is described by Liouvillian superoperators that map a density operator to another density operator~\cite{Nielsen-textbook, Breuer-textbook, Rivas-textbook}.
In addition, non-Hermitian Hamiltonians effectively capture the open quantum dynamics of individual quantum trajectories subject to measurements or dissipative processes~\cite{Carmichael-textbook, Plenio-review, Daley-review}, as well as the open classical dynamics~\cite{Bender-review, Konotop-review, Christodoulides-review}.
Researchers have discovered unique phenomena that lack analogs in closed quantum systems and hence are intrinsic to open quantum systems.
Prime recent examples include the entanglement phase transitions induced by the interplay of the coherent dynamics and measurement~\cite{Potter-Vasseur-2022, Fisher-review}, as well as non-Hermitian topological phases~\cite{Bergholtz-review, Okuma-Sato-review}.

Moreover, characterization of chaos in open quantum systems has also attracted growing interest~\cite{Grobe-88, *Grobe-89, Xu-19, Hamazaki-19, Denisov-19, Can-19PRL, *Can-19JPhysA, Hamazaki-20, Akemann-19, Sa-20, Wang-20, Xu-21, GarciaGarcia-22PRL, *GarciaGarcia-22PRD, JiachenLi-21, Tarnowski-21, Cornelius-22, GarciaGarcia-22PRX, Sa-22-SYK, Kulkarni-22-SYK, GJ-23, *GJ-23-ETH, Xiao-22, Shivam-22, Ghosh-22, Matsoukas-Roubeas-23, Sa-22, Kawabata-22}.
Such dissipative quantum chaos is relevant to the development of thermodynamics and statistical mechanics in open quantum systems.
As prototypes of open quantum chaotic systems, researchers have proposed random Lindbladians~\cite{Xu-19, Denisov-19, Can-19PRL, *Can-19JPhysA, Wang-20}, as well as dissipative extensions of the Sachdev-Ye-Kitaev model~\cite{Xu-21, GarciaGarcia-22PRL, *GarciaGarcia-22PRD, GarciaGarcia-22PRX, Sa-22-SYK, Kulkarni-22-SYK, Kawabata-22}.
Analogous to closed quantum systems, several numerical calculations have shown that the statistics of complex-valued eigenvalues in nonintegrable open quantum systems follow the statistics of non-Hermitian random matrices, while those in integrable counterparts follow the Poisson statistics for complex numbers.
Notably, non-Hermiticity changes the nature of symmetry, and the spectral statistics of non-Hermitian random matrices are no longer described by the 10-fold symmetry classification, instead universally conforming to the 38-fold symmetry classification~\cite{Bernard-LeClair-02, KSUS-19}.
These 38-fold symmetry classes are also relevant to the classification of the Anderson transitions~\cite{Hatano-Nelson-96, *Hatano-Nelson-97, Efetov-97, Feinberg-97, Brouwer-97, Longhi-19, Tzortzakakis-20, Huang-20, KR-21, Luo-21L,* Luo-21B, Luo-22R, Liu-Fulga-21, Ghosh-23} and topological phases~\cite{Gong-18, KSUS-19, Zhou-Lee-19, Bergholtz-review, Okuma-Sato-review} in non-Hermitian systems.

In this work, we investigate the spectral statistics of singular values, instead of complex eigenvalues, in non-Hermitian random matrices.
For a given matrix $H$, its singular values are defined as the square roots of eigenvalues of $H^{\dag} H$ or $HH^{\dag}$.
When $H$ respects Hermiticity (i.e., $H^{\dag} = H$), singular values reduce to the absolute values of eigenvalues and hence contain essentially the same information as eigenvalues.
However, in the case of non-Hermitian matrices $H$ (i.e., $H^{\dag} \neq H$), singular values play a distinct role from eigenvalues and find broad applications in diverse areas of science.
For example, singular values represent the degree of information associated with each mode and thus form the basis for principal component analysis and various tensor-network techniques.
This also implies that if $H$ is interpreted as a generator of the open quantum dynamics, its singular values quantify the amount of information in the quantum operation.
The singular-value decomposition is also utilized to capture the 
topological phenomena and
bulk-boundary correspondence in non-Hermitian systems~\cite{Porras-19, *Ramos-22, Herviou-19, Wanjura-20, *Brunelli-22}.
Additionally, if $H$ is interpreted as a reflection or transmission matrix, the squared singular values give the reflection or transmission probability, respectively~\cite{Datta-textbook, Imry-textbook}.
The fluctuations of singular values from sample to sample, or the suppression thereof, physically manifest themselves in the universal conductance fluctuations~\cite{Lee-85, *Lee-87, Altshuler-85, Imry-86, Altshuler-86}.

Here, we demonstrate that the singular-value statistics provide an effective diagnostic tool for understanding chaotic behavior in open quantum systems.
We establish that singular values of open quantum many-body systems, including Lindbladians and non-Hermitian Hamiltonians, follow the random-matrix statistics in the corresponding symmetry classes.
Employing the Hermitization technique~\cite{Feinberg-97, Gong-18, KSUS-19, Luo-22R}, we completely classify the singular-value statistics of non-Hermitian random matrices across all the 38 symmetry classes, as summarized in Tables~\ref{tab: complex AZ}-\ref{tab: real AZ + SLS}.
Importantly, we reveal that the singular-value statistics exhibit unique characteristics and serve as a clearer indicator of symmetry compared to the complex-eigenvalue statistics.
In contrast to Hermitian random matrices, the level repulsion of complex eigenvalues in non-Hermitian random matrices is universally cubic, regardless of symmetry~\cite{Grobe-89}.
Additionally, the correlations of complex eigenvalues in the bulk 
of the spectra
depend solely on time-reversal symmetry$^{\dag}$, and the other symmetries are irrelevant~\cite{Hamazaki-20, Xiao-22}.
By contrast, we elucidate that all the symmetries influence the bulk spectral correlations of singular values even in non-Hermitian random matrices.
Furthermore, we illustrate that these distinctions can be clearly discerned through the random-matrix indices $\left( \alpha, \beta \right)$, akin to Hermitian random matrices.
Consequently, these results show the significance of singular-value statistics as a powerful practical measure for characterizing dissipative quantum chaos and its absence, thereby facilitating a deeper understanding of statistical mechanics in open quantum systems.

The rest of this work is organized as follows.
In Sec.~\ref{sec: symmetry}, we describe symmetry of non-Hermitian matrices.
In Sec.~\ref{sec: singular-value statistics}, we study the singular-value statistics of non-Hermitian random matrices (Table~\ref{tab: NH RMT} and Fig.~\ref{fig: NH RMT}).
Through Hermitization, we completely classify the singular-value statistics of non-Hermitian random matrices for all the 38 symmetry classes (Tables~\ref{tab: complex AZ}-\ref{tab: real AZ + SLS}). 
In Sec.~\ref{sec: dissipative quantum chaos}, we demonstrate that the singular-value statistics of physical Lindbladians and non-Hermitian Hamiltonians  coincide with those of non-Hermitian random matrices, characterizing the dissipative quantum chaos.
In Sec.~\ref{sec: conclusion}, we conclude this work.
In Appendix~\ref{asec: Hermitian}, we discuss the singular-value statistics of Hermitian random matrices (Table~\ref{tab: singular value - Hermitian}).
In Appendix~\ref{asec: Wigner surmise}, we analytically derive the singular-value statistics of small random matrices for all the symmetry classes.
In Appendix~\ref{asec: normal}, we discuss the singular-value statistics of normal random matrices.
In Appendix~\ref{asec: reflection}, we develop the symmetry classification of non-Hermitian reflection matrices.

\section{Symmetry of non-Hermitian random matrices}
    \label{sec: symmetry}

We begin with the symmetry classification of Hermitian random matrices.
In this work, we focus on square matrices to consider physical applications to quantum chaos.
In general, Hermitian matrices $H$ are classified according to the two types of antiunitary symmetry, time-reversal symmetry
\begin{equation}
    \mathcal{T} H^{*} \mathcal{T}^{-1} = H,\quad \mathcal{T}\mathcal{T}^{*} = \pm 1, 
        \label{eq: TRS} 
\end{equation}
and particle-hole symmetry
\begin{equation}
    \mathcal{C} H^{T} \mathcal{C}^{-1} = - H,\quad \mathcal{C}\mathcal{C}^{*} = \pm 1, 
        \label{eq: PHS}
\end{equation}
with unitary matrices $\mathcal{T}$ and $\mathcal{C}$.
As a combination of time-reversal symmetry and particle-hole symmetry, we can also introduce chiral symmetry, or equivalently sublattice symmetry, by
\begin{equation}
    \mathcal{S} H \mathcal{S}^{-1} = - H,\quad \mathcal{S}^2 = +1, 
        \label{eq: SLS}
\end{equation}
with a unitary matrix $\mathcal{S}$.
These two antiunitary symmetries and one unitary symmetry form the 10-fold Altland-Zirnbauer (AZ) symmetry classification (see also Table~\ref{tab: singular value - Hermitian} in Appendix~\ref{asec: Hermitian})~\cite{AZ-97}.
Classes A, AI, and AII are the 3-fold standard (Wigner-Dyson) classes~\cite{Wigner-51, *Wigner-58, Dyson-62}, classes AIII, BDI, and CII are the 3-fold chiral classes~\cite{Gade-91, *Gade-93, Verbaarschot-94, *Verbaarschot-00-review}, and classes D, DIII, C, and CI are the 4-fold Bogoliubov-de Gennes (BdG) classes~\cite{AZ-97}.
These AZ symmetry classes yield the 10-fold classification of Hermitian random matrices~\cite{AZ-97, Haake-textbook}, as well as the Anderson transitions~\cite{Beenakker-review-97, *Beenakker-review-15, Evers-review} and topological insulators and superconductors~\cite{Schnyder-08, *Ryu-10, Kitaev-09, HK-review, QZ-review, CTSR-review}.

The nature of symmetry changes in non-Hermitian matrices, leading to the 38-fold classification~\cite{Bernard-LeClair-02, KSUS-19}.
First, the two types of antiunitary symmetry in Eqs.~(\ref{eq: TRS}) and (\ref{eq: PHS}) remain to be time-reversal symmetry and particle-hole symmetry even for non-Hermitian matrices $H$, respectively.
In addition to these symmetries, we can also introduce two additional antiunitary symmetries by
\begin{equation}
    \mathcal{T} H^{T} \mathcal{T}^{-1} = H,\quad \mathcal{T}\mathcal{T}^{*} = \pm 1, 
        \label{eq: TRS-dag}
\end{equation}
and
\begin{equation}
    \mathcal{C} H^{*} \mathcal{C}^{-1} = - H,\quad \mathcal{C}\mathcal{C}^{*} = \pm 1, 
        \label{eq: PHS-dag}
\end{equation}
with unitary matrices $\mathcal{T}$ and $\mathcal{C}$.
These symmetries are respectively denoted by time-reversal symmetry$^{\dag}$ and particle-hole symmetry$^{\dag}$ since they are obtained by additional Hermitian conjugation to time-reversal symmetry in Eq.~(\ref{eq: TRS}) and particle-hole symmetry in Eq.~(\ref{eq: PHS}), respectively~\cite{KSUS-19}.
While time-reversal symmetry in Eq.~(\ref{eq: TRS}) and time-reversal symmetry$^{\dag}$ in Eq.~(\ref{eq: TRS-dag}) are equivalent for Hermitian matrices, this is not the case for non-Hermitian matrices.

The unitary symmetry in Eq.~(\ref{eq: SLS}) is still symmetry and is denoted by sublattice symmetry for non-Hermitian matrices.
Similarly to the antiunitary symmetries, we can also consider the Hermitian conjugate of Eq.~(\ref{eq: SLS}) by
\begin{equation}
    \mathcal{S} H^{\dag} \mathcal{S}^{-1} = - H,\quad \mathcal{S}^2 = +1, 
        \label{eq: CS}
\end{equation}
with a unitary matrix $\mathcal{S}$.
In contrast to Eq.~(\ref{eq: SLS}), unitary symmetry in Eq.~(\ref{eq: CS}) is called chiral symmetry for non-Hermitian matrices.
Here, while chiral symmetry and sublattice symmetry are equivalent to each other for Hermitian matrices, they are different for non-Hermitian matrices.
The combination of time-reversal symmetry in Eq.~(\ref{eq: TRS}) and particle-hole symmetry in Eq.~(\ref{eq: PHS}), as well as the combination of time-reversal symmetry$^{\dag}$ in Eq.~(\ref{eq: TRS-dag}) and particle-hole symmetry$^{\dag}$ in Eq.~(\ref{eq: PHS-dag}), yields chiral symmetry in Eq.~(\ref{eq: CS});
on the other hand, the combination of time-reversal symmetry and particle-hole symmetry$^{\dag}$, as well as the combination of time-reversal symmetry$^{\dag}$ and particle-hole symmetry, yields sublattice symmetry in Eq.~(\ref{eq: SLS}).
Chiral symmetry in Eq.~(\ref{eq: CS}) is also equivalent to pseudo-(anti-)Hermiticity~\cite{Mostafazadeh-02-1, *Mostafazadeh-02-2}.

Different symmetries lead to different universality classes, resulting in the 38-fold classification of non-Hermitian random matrices.
Specifically, time-reversal symmetry$^{\dag}$ in Eq.~(\ref{eq: TRS-dag}) changes the complex-eigenvalue statistics of non-Hermitian random matrices in the bulk of the spectra~\cite{Hamazaki-20} while the other symmetries change the complex-eigenvalue statistics at the symmetric lines or point~\cite{Xiao-22}.
Moreover, the 38-fold classification characterizes the universality classes of the Anderson transitions~\cite{KR-21, Luo-21L,* Luo-21B, Luo-22R} and topological phases~\cite{KSUS-19, Zhou-Lee-19} of non-Hermitian systems.
In this work, we below show that symmetry determines the universal statistics of singular values in non-Hermitian random matrices and comprehensively classify them for all the 38 symmetry classes.

\section{Singular-value statistics of non-Hermitian random matrices}
    \label{sec: singular-value statistics}

The complex-eigenvalue statistics of non-Hermitian random matrices have recently been investigated as a diagnostic of dissipative quantum chaos.
Here, we study the spectral statistics of singular values in non-Hermitian random matrices.
While eigenvalues and singular values contain essentially the same information in Hermitian matrices, they exhibit distinct behavior in non-Hermitian matrices.
We demonstrate that the singular-value statistics exhibit the behavior that distinguishes them from the complex-eigenvalue statistics and serve as an effective measure of dissipative quantum chaos.
While we focus on non-Hermitian random matrices from the Gaussian ensemble 
$p \left( H \right) \propto e^{-\mathrm{tr}\,(H^{\dag}H)/2}$
for clarity, our results should be independent of the specific details of the ensemble and universally determined by symmetry alone.
In fact, we below show that the random-matrix statistics appear even in open quantum chaotic systems, suggesting the universality of the singular-value statistics (see Sec.~\ref{sec: dissipative quantum chaos} for details).

\begin{figure}[t]
\centering
\includegraphics[width=\linewidth]{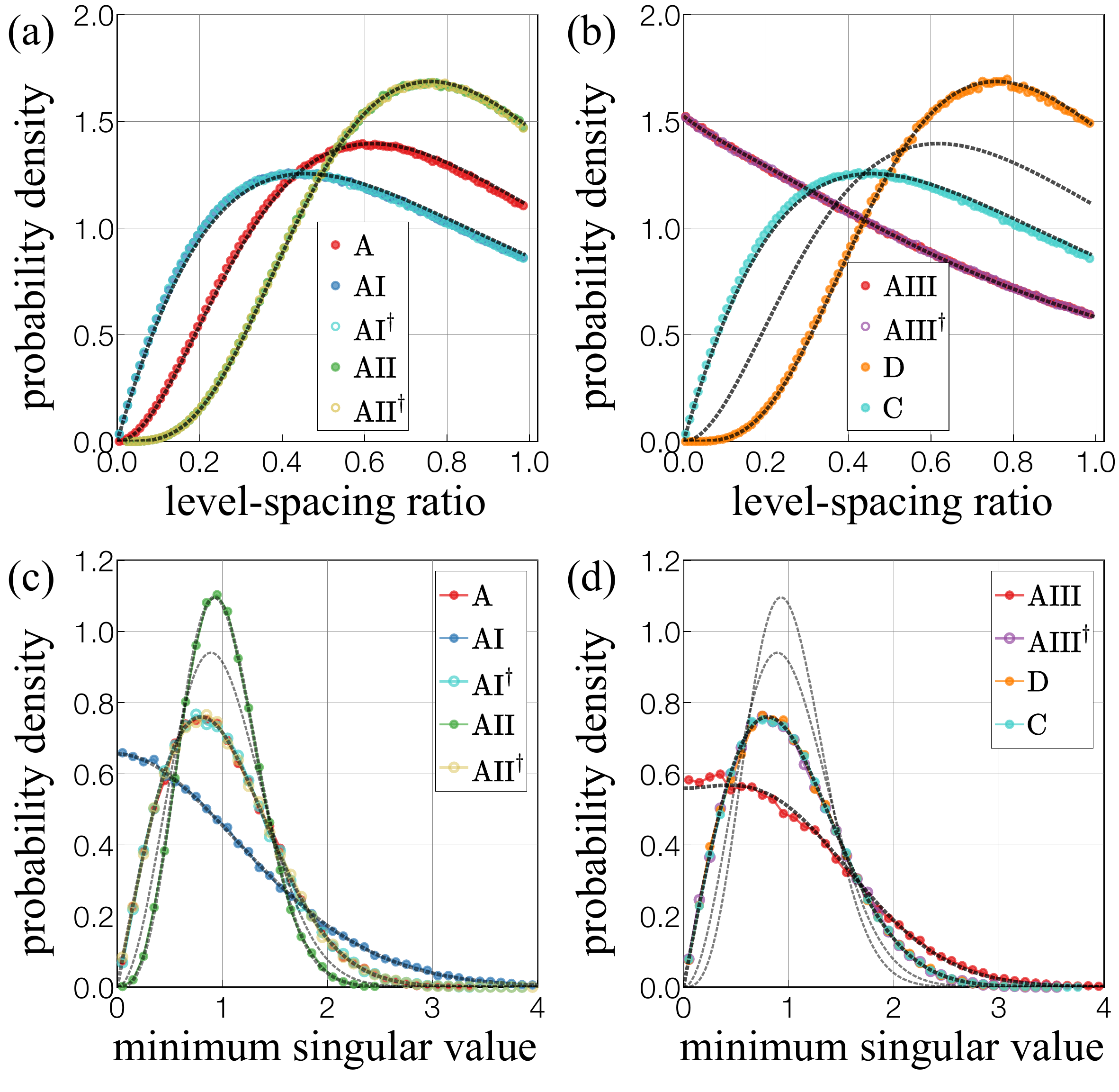} 
\caption{Singular-value statistics of non-Hermitian random matrices.
All the results are averaged over (a, b)~$10^4$ and (c, d)~$5 \times 10^4$ realizations of $10^3 \times 10^3$ matrices in the Gaussian ensemble.
(a, b)~Level-spacing-ratio distributions of singular values in classes (a)~A, AI, AI$^{\dag}$, AII, AII$^{\dag}$, (b)~AIII, AIII$^{\dag}$, D, and C.
The black dashed curves are the analytical results for small random matrices [Eqs.~(\ref{aeq: Wigner-pr-chiral&BdG}) and (\ref{aeq: Wigner-pr-standard})].
Classes AI, AI$^{\dag}$, and C (classes AII, AII$^{\dag}$, and D) follow the same distribution characterized by $\beta = 1$ ($\beta = 4$).
Classes AIII and AIII$^{\dag}$ follow the same distribution.
(c, d)~Distributions of the minimum singular value $s_{\rm min}$ in classes (c)~A, AI, AI$^{\dag}$, AII, AII$^{\dag}$, (d)~AIII, AIII$^{\dag}$, D, and C.
The probability distribution functions are normalized such that their averages $\braket{s_{\rm min}}$ are $1$.
The black dashed curves are the analytical results for small random matrices [Eqs.~(\ref{aeq: s_min_dis_chiral}), (\ref{aeq: s_min_dis_BdG}), and (\ref{aeq: s_min_dis_standard})].
Classes A, AI$^{\dag}$, AII$^{\dag}$, AIII$^{\dag}$, D, and C follow the same distribution characterized by $\alpha = 1$.}
No symmetry classes considered here and listed in Table~\ref{tab: NH RMT} follow Eq.~(\ref{aeq: s_min_dis_BdG}) with $\alpha = 2$.
	\label{fig: NH RMT}
\end{figure}

In Sec.~\ref{subsec: singular-value statistics}, we first provide the singular-value statistics for the 9 symmetry classes where no symmetry or only one symmetry is present (Table~\ref{tab: NH RMT} and Fig.~\ref{fig: NH RMT}).
In Sec.~\ref{subsec: Hermitization}, we reduce the singular-value statistics of non-Hermitian random matrices to those of Hermitian random matrices in the corresponding symmetry classes, using the Hermitization technique.
In Sec.~\ref{subsec: classification}, we provide the complete classification of the singular-value statistics of non-Hermitian random matrices for all the 38 symmetry classes (Tables~\ref{tab: complex AZ}-\ref{tab: real AZ + SLS}).

\begin{table*}[t]
	\centering
	\caption{Singular-value statistics of non-Hermitian random matrices.
    Class A is characterized by no symmetry;
    classes AI and AII (AI$^{\dag}$ and AII$^{\dag}$) are characterized by time-reversal symmetry (time-reversal symmetry$^{\dag}$);
    class AIII (AIII$^{\dag}$) is characterized by chiral (sublattice) symmetry; 
    classes D and C are characterized by particle-hole symmetry.
    The number $+1$ or $-1$ in each parentheses specifies the sign of each antiunitary symmetry.
    In classes AII, AII$^{\dag}$, and D, all the singular values are two-fold degenerate.
    The corresponding Hermitized symmetry classes are shown in the column ``Hermitization''.
    There, ``AIII $\times$ AIII'' means that the Hermitized matrix is decomposed into two independent Hermitian matrices in class AIII (see Sec.~\ref{subsec: Hermitization} for details).
    The analytical results of the average singular-value-spacing ratios $\braket{r}$ and the average normalized variances $\braket{s_{\rm min}^2}/\braket{s_{\rm min}}^2$ of the minimum singular value for small random matrices are shown.}
	\label{tab: NH RMT}
     \begin{tabular}{ccccc} \hline \hline
    ~~Class~~ & ~~Symmetry~~ & ~~Hermitization~~ & ~~$\braket{r}$~~ & ~~$\braket{s_{\rm min}^2}/\braket{s_{\rm min}}^2$~~ \\ \hline
    A & ~~No~~ & AIII & $0.6026$ & $1.2732$ \\
    AI & ~~Time-reversal symmetry ($+1$)~~ & BDI & $0.5358$ & $1.6018$ \\
    AI$^{\dag}$ & ~~Time-reversal symmetry$^{\dag}$ ($+1$)~~ & CI & $0.5358$ & $1.2732$ \\
    AII & ~~Time-reversal symmetry ($-1$)~~ & CII & $0.6761$ & $1.1291$ \\
    AII$^{\dag}$ & ~~Time-reversal symmetry$^{\dag}$ ($-1$)~~ & DIII & $0.6761$ & $1.2732$ \\ \hline
    AIII & ~~Chiral symmetry~~ & A & $0.4206$ & $1.4786$ \\
    AIII$^{\dag}$ & ~~Sublattice symmetry~~ & AIII $\times$ AIII & $0.4206$ & $1.2732$ \\ \hline
    D & ~~Particle-hole symmetry ($+1$)~~ & DIII & $0.6761$ & $1.2732$ \\
    C & ~~Particle-hole symmetry ($-1$)~~ & CI & $0.5358$ & $1.2732$ \\ \hline \hline
  \end{tabular}
\end{table*}

\subsection{Singular-value statistics}
    \label{subsec: singular-value statistics}

We first focus on the 9 symmetry classes where no symmetry or only one symmetry is present (Table~\ref{tab: NH RMT}) and provide the complete classification for all the 38 symmetry classes shortly in Sec.~\ref{subsec: classification} (Tables~\ref{tab: complex AZ}-\ref{tab: real AZ + SLS}).
Singular values of a matrix $H$ are the square roots of eigenvalues of $H^{\dag} H$ or $HH^{\dag}$.
As a clear difference from eigenvalues, singular values are always real and nonnegative even in non-Hermitian matrices.
Then, we calculate the distributions of singular-value-spacing ratios $r_n$ [Fig.~\ref{fig: NH RMT}\,(a, b)], in a similar manner to the spacing ratios of real eigenvalues~\cite{Oganesyan-Huse-07, Atas-13}.
Specifically, $r_n$ is defined as
\begin{align}
    r_{n} \coloneqq \min \left( \frac{s_{n+1} - s_{n}}{s_{n} - s_{n-1}}, \frac{s_{n} - s_{n-1}}{s_{n+1} - s_{n}}\right),\quad 0 \leq r_{n} \leq 1,
        \label{eq: spacing ratio}
\end{align}
for an ordered set of singular values $s_{n}$'s.
In classes AII, AII$^{\dag}$, and D, all the singular values are two-fold degenerate, and we calculate $r_n$ by identifying the two degenerate singular values with the one singular value.
While we can also study the statistical distributions of singular-value spacing $s_{n+1} - s_{n}$ (see also Appendix~\ref{asec: Wigner surmise}), the statistical distributions of $r_n$ are independent of the local density of singular values and hence do not require the unfolding of singular values, which facilitates the calculations for physical open quantum systems.

The singular-value-spacing ratios exhibit the three different distributions in the presence of time-reversal symmetry in Eq.~(\ref{eq: TRS}) or time-reversal symmetry$^{\dag}$ in Eq.~(\ref{eq: TRS-dag}).
Here, non-Hermitian matrices without symmetry belong to class A, and those with Eq.~(\ref{eq: TRS}) [Eq.~(\ref{eq: TRS-dag})] belong to classes AI and AII (AI$^{\dag}$ and AII$^{\dag}$)~\cite{KSUS-19}.
Notably, time-reversal symmetry changes the level-spacing distributions even for arbitrary singular values.
By contrast, while time-reversal symmetry$^{\dag}$ changes the bulk statistics of complex eigenvalues~\cite{Hamazaki-19}, time-reversal symmetry changes the eigenvalue statistics only around the real axis~\cite{Ginibre-65, Girko-85, Mehta-textbook, Xiao-22, 2022arXiv221116223B, *2023arXiv230105022B}.
Thus, the singular-value statistics give a clearer measure of time-reversal symmetry in non-Hermitian random matrices.
It is also notable that all the level-spacing distributions of complex eigenvalues exhibit the cubic level repulsion~\cite{Grobe-89, Hamazaki-20, Sa-20} in contrast to Hermitian random matrices.
Conversely, the singular-value-spacing distributions exhibit the 3-fold level repulsion even in non-Hermitian random matrices,
\begin{align}
p_{\rm r} \propto r^{\beta}\quad \left( 0 \leq r \ll 1 \right)
    \label{eq: beta}
\end{align}
for 
\begin{align}
    \beta = \begin{cases}
        1 & ( \text{classes~AI~and~AI}^{\dag} ); \\
        2 & ( \text{class~A} ); \\
        4 & ( \text{classes~AII~and~AII}^{\dag} ).
    \end{cases}
\end{align}
All the other symmetries also change the level-spacing-ratio distributions of singular values [Fig.~\ref{fig: NH RMT}\,(b)].
Remarkably, in classes AIII and AIII$^{\dag}$, where chiral and sublattice symmetries in Eqs.~(\ref{eq: CS}) and (\ref{eq: SLS}) are respected by definition, the probability density is nonvanishing even for small level-spacing ratio $0 \leq r \ll 1$, which implies the weaker level repulsion between singular values.
Still, the level-spacing-ratio distributions deviate from the Poisson statistics 
\begin{align}
    p_{\rm r} \left( r \right) = \frac{2}{\left( 1+r \right)^2}
        \label{eq: pr-Poisson}
\end{align}
for uncorrelated singular values.
We elucidate these unconventional level-spacing-ratio distributions through Hermitization in Sec.~\ref{subsec: Hermitization}.

While some symmetry classes (e.g., classes AI and AI$^{\dag}$) exhibit the same level-spacing distributions, they can be distinguished by the distributions of the minimum singular value $s_{\rm min}$ [Fig.~\ref{fig: NH RMT}\,(c, d)].
Here, in contrast to the spacing ratio $r$, the minimum singular value $s_{\rm min}$ depends on 
the
normalization of random matrices;
in our numerical calculations in Fig.~\ref{fig: NH RMT}\,(c, d), we normalize $s_{\rm min}$ such that its average is unity (i.e., $\braket{s_{\rm min}} = 1$).
For example, non-Hermitian random matrices in classes AI and AI$^{\dag}$ follow the same level-spacing distributions but exhibit clearly different distributions of the minimum singular value.
The former is quantified by the average singular-value-spacing ratio $\braket{r}$ while the latter by the average normalized variances $\braket{s_{\rm min}^2}/\braket{s_{\rm min}}^2$ of the minimum singular value, as summarized in Table~\ref{tab: NH RMT}.
Additionally, each symmetry class can be distinguished between by the power-law decay of the probability distribution $p_{\rm min}$ for small $s_{\rm min}$, 
\begin{align}
    p_{\rm min} \propto s_{\rm min}^{\alpha} \quad \left( 0 \leq s_{\rm min} \ll 1 \right)
        \label{eq: alpha}
\end{align}
with 
\begin{align}
    \alpha = \begin{cases}
        0 & ( \text{class~AI} ); \\
        1 & ( \text{classes~A,~AI}^{\dag}, \text{AII}^{\dag}, \text{AIII}^{\dag}, \text{D,~and~C} ); \\
        3 & ( \text{class~AII} ).
    \end{cases}
\end{align}
We also note that class AIII is characterized by unique singular-value statistics, as discussed shortly.
If matrices additionally respect Hermiticity or normality,
they exhibit different singular-value statistics (see Appendices~\ref{asec: Hermitian} and \ref{asec: normal} for details).

\subsection{Hermitization}
    \label{subsec: Hermitization}

We demonstrate that the singular-value statistics are captured by Hermitization. 
Previously, the Hermitization technique was applied to the complex-eigenvalue statistics~\cite{Feinberg-97}.
It is also relevant to the classification of the Anderson transitions~\cite{Luo-22R} and topological phases~\cite{Gong-18, KSUS-19} in non-Hermitian systems.
For a given non-Hermitian matrix $H$, we introduce a Hermitized matrix
\begin{align}
    \tilde{H} \coloneqq \begin{pmatrix}
        0 & H \\
        H^{\dag} & 0
    \end{pmatrix},\quad \tilde{H}^{2} = \begin{pmatrix}
        HH^{\dag} & 0 \\
        0 & H^{\dag} H
    \end{pmatrix}.
\end{align}
Singular values of $\tilde{H}$ are given as eigenvalues of $\sqrt{HH^{\dag}}$ and $\sqrt{H^{\dag}H}$, and hence coincide with singular values of $H$.
Thus, the singular-value statistics of non-Hermitian matrices $H$ reduce to those of the corresponding Hermitian matrix $\tilde{H}$.
We provide the singular-value statistics of Hermitian random matrices in Appendix~\ref{asec: Hermitian} (Table~\ref{tab: singular value - Hermitian}).
Importantly, Hermitized matrices $\tilde{H}$ respect additional chiral symmetry by construction,
\begin{align}
    \sigma_z \tilde{H} \sigma_z = - \tilde{H},\quad 
    \sigma_z \coloneqq \begin{pmatrix}
        1 & 0 \\
        0 & -1
    \end{pmatrix},
\end{align}
which changes the relevant symmetry classes.
For example, for non-Hermitian matrices that respect only time-reversal symmetry in Eq.~(\ref{eq: TRS}) (i.e., classes AI or AII), the Hermitized matrices belong to classes BDI or CII.
We also summarize the correspondence of symmetry classes in Table~\ref{tab: NH RMT}, as well as Tables~\ref{tab: complex AZ}-\ref{tab: real AZ + SLS}.

Through Hermitization, the singular-value statistics of non-Hermitian random matrices $H$ with time-reversal symmetry in Eq.~(\ref{eq: TRS}), particle-hole symmetry in Eq.~(\ref{eq: PHS}), time-reversal symmetry$^{\dag}$ in Eq.~(\ref{eq: TRS-dag}), or particle-hole symmetry$^{\dag}$ in Eq.~(\ref{eq: PHS-dag}) reduce to the real-eigenvalue statistics of Hermitized matrices $\tilde{H}$ in the chiral~\cite{Edelman-88, Forrester-93, Nishigaki-98, *Damgaard-01} and BdG~\cite{AZ-97, Sun-20} classes.
In fact, this correspondence is consistent with our numerical calculations of the singular-value statistics (compare Table~\ref{tab: NH RMT} and Fig.~\ref{fig: NH RMT} with Table~\ref{tab: singular value - Hermitian} and Fig.~\ref{fig: Hermitian RMT}, respectively).
The joint probability distribution functions of singular values $s_i$'s are given as
\begin{align}
    \rho \left( \{ s_i \} \right) \propto \prod_{i} s_{i}^{\alpha} \prod_{i<j} \left| s_i - s_j \right|^{\beta} e^{- \sum_{i} s_i^2}
        \label{eq: jpdf}
\end{align}
with the random-matrix indices $\alpha$ in Eq.~(\ref{eq: alpha}) and $\beta$ in Eq.~(\ref{eq: beta}).

By contrast, in the presence of chiral symmetry in Eq.~(\ref{eq: CS}), the Hermitized matrices $\tilde{H}$ satisfy
\begin{align}
    \tilde{U}^{\dag} \tilde{H} \tilde{U} = \begin{pmatrix}
        \ii H \mathcal{S} & 0 \\
        0 & - \ii H \mathcal{S}
    \end{pmatrix},\quad \tilde{U} \coloneqq \frac{1}{\sqrt{2}} \begin{pmatrix}
        1 & -\ii \\
        \ii \mathcal{S} & -\mathcal{S}
    \end{pmatrix},
        \label{eq: chiral-iHS}
\end{align}
and hence the singular values of $\tilde{H}$ further reduce to the singular values of the Hermitian matrices $\ii H \mathcal{S}$ and $- \ii H \mathcal{S}$ without any symmetry (i.e., class A).
When the non-Hermitian matrix $H$ is taken from the Gaussian ensemble, the Hermitian matrices $\ii H \mathcal{S}$ and $- \ii H \mathcal{S}$ also belong to the Gaussian ensemble.
Since singular values of Hermitian matrices are given as the absolute values of their eigenvalues, the singular-value statistics without symmetry reduce to the real-eigenvalue statistics of two independent ensembles one of which corresponds to positive eigenvalues and the other of which corresponds to negative eigenvalues.
There, uncorrelated levels can be close to each other, and the level repulsion is significantly weakened, which is the origin of $p_{\rm r} \left( r=0 \right) \neq 0$ in Fig.~\ref{fig: NH RMT}\,(b).
Still, a part of the level repulsion survives, resulting in the deviation from the Poisson statistics.

Furthermore, non-Hermitian matrices $H$ with sublattice symmetry in Eq.~(\ref{eq: SLS}) can be generally expressed as
\begin{align}
    H = \begin{pmatrix}
        0 & h_1 \\
        h_2 & 0 
    \end{pmatrix}
        \label{eq: SLS-block}
\end{align}
with two non-Hermitian matrices $h_1$ and $h_2$, where we assume $\mathcal{S} = \sigma_z$.
Hence, the singular values of $H$ consist of the singular values of the two independent non-Hermitian matrices $h_1$ and $h_2$.
Consequently, the spacing distributions in the bulk
of the singular-value spectra
again reduce to those of Hermitian matrices without symmetry.
On the other hand, the minimum singular value of $H$ reduces to the smaller one of the minimum singular values of $h_1$ and $h_2$, the distribution of which obeys that of Hermitian matrices with chiral symmetry (a subtle difference can arise in generic cases; see Sec.~\ref{subsec: classification} for details).
In the absence of chiral or sublattice symmetry, positive and negative eigenvalues of Hermitized matrices $\tilde{H}$ are correlated owing to symmetry, leading to the level repulsion between singular values.

\subsection{Classification}
    \label{subsec: classification}

Based on Hermitization, we completely classify the singular-value statistics of non-Hermitian random matrices for all the 38 symmetry classes in Tables~\ref{tab: complex AZ}-\ref{tab: real AZ + SLS}, 9 of which are summarized also in Table~\ref{tab: NH RMT}.
In a similar manner to the 10-fold AZ symmetry classes for Hermitian matrices, time-reversal symmetry in Eq.~(\ref{eq: TRS}), particle-hole symmetry in Eq.~(\ref{eq: PHS}), and chiral symmetry in Eq.~(\ref{eq: CS}) form the 10-fold symmetry classes for non-Hermitian matrices (Tables~\ref{tab: complex AZ} and \ref{tab: real AZ}).
Furthermore, time-reversal symmetry$^{\dag}$ in Eq.~(\ref{eq: TRS-dag}), particle-hole symmetry$^{\dag}$ in Eq.~(\ref{eq: PHS-dag}), and chiral symmetry in Eq.~(\ref{eq: CS}) form additional 10-fold symmetry classes, which are the AZ$^{\dag}$ symmetry classes for non-Hermitian matrices (Tables~\ref{tab: complex AZ} and \ref{tab: real AZ-dag}).
Taking sublattice symmetry into consideration as additional symmetry (Tables~\ref{tab: complex AZ} and \ref{tab: real AZ + SLS}), we have the 38-fold symmetry classification for non-Hermitian matrices~\cite{KSUS-19}.
Some symmetry classes in Tables~\ref{tab: complex AZ}-\ref{tab: real AZ + SLS} give the equivalent symmetry classes, which are not double counted in the 38-fold symmetry classification.
For example, time-reversal symmetry in Eq.~(\ref{eq: TRS}) and particle-hole symmetry$^{\dag}$ in Eq.~(\ref{eq: PHS-dag}) lead to essentially the same universal spectral statistics of both eigenvalues and singular values~\cite{Kawabata-19}.
Consequently, classes AI and AII are equivalent to classes D$^{\dag}$ and C$^{\dag}$, respectively (Table~\ref{tab: real AZ}).

In these classification tables, we provide the random-matrix indices $\beta$ for the singular-value statistics in the bulk of the spectra [i.e., Eq.~(\ref{eq: beta})] and $\alpha$ for the statistics of the minimum singular value [Eq.~(\ref{eq: alpha})].
On the other hand, the symmetry classes for which the corresponding Hermitized matrices belong to the standard classes (i.e., classes A, AI, and AII) are not characterized by these random-matrix indices $\left( \alpha, \beta \right)$, as discussed above.
In the symmetry classes for which the Hermitized matrices respect additional unitary symmetry and hence are block diagonalized, the singular-value statistics in the bulk of the spectra are no longer described by $\beta$, 
whereas the statistics of the minimum singular value can still be described by $\alpha$.
Furthermore, we analytically derive the singular-value statistics of small random matrices in Appendix~\ref{asec: Wigner surmise}.
These small-$N$ analytical results well describe the large-$N$ results, in the similar spirit to the Wigner surmise~\cite{Wigner-51, *Wigner-58, Dyson-62, Atas-13}.
This also contrasts with the complex-eigenvalue statistics of non-Hermitian random matrices, where a significant deviation arises between the large-$N$ and small-$N$ results~\cite{Hamazaki-20,Xiao-22}.

In the presence of chiral symmetry in Eq.~(\ref{eq: CS}), singular values of a non-Hermitian matrix $H$ generally reduce to singular values of Hermitian matrices $\ii H \mathcal{S}$ and $- \ii H \mathcal{S}$ with certain symmetry, as shown in Eq.~(\ref{eq: chiral-iHS}).
It should be noted that the spectral supports of $\ii H \mathcal{S}$ and $- \ii H \mathcal{S}$ can differ from each other for generic random matrices, as well as physical models such as Lindbladians and non-Hermitian Hamiltonians (see Sec.~\ref{sec: dissipative quantum chaos} below). 
The two spectra may not overlap when all the eigenvalues of $\ii H \mathcal{S}$ ($- \ii H \mathcal{S}$) are positive (negative). 
In such a case, the level repulsion between eigenvalues is manifested also in the singular-value statistics, which contrasts with the weaker level repulsion discussed in Sec.~\ref{subsec: Hermitization}.
Conversely, for the non-Hermitian matrix $H$ from the Gaussian ensemble, the two Hermitian matrices $\ii H \mathcal{S}$ and $- \ii H \mathcal{S}$ also belong to the Gaussian ensemble and share the same spectral support. 
Similarly, to obtain the universal results in the presence of chiral symmetry, the spectrum of the Hermitian matrices $\pm \ii H \mathcal{S}$ should be chosen to be statistically symmetric with respect to zero.

Furthermore, in the presence of sublattice symmetry in Eq.~(\ref{eq: SLS}), singular values of a non-Hermitian matrix $H$ generally decompose into singular values of two independent non-Hermitian matrices $h_1$ and $h_2$ in Eq.~(\ref{eq: SLS-block}).
In particular, the minimum singular value of $H$ reduces to the smaller one of the minimum singular values of $h_1$ and $h_2$.
In the sole presence of sublattice symmetry (i.e., class AIII$^{\dag}$ in Table~\ref{tab: complex AZ}), the distribution of the minimum singular value in $H$ coincides with that in $h_1$ or $h_2$.
However, this is not necessarily the case, in general.
For example, in the additional presence of time-reversal symmetry that commutes with sublattice symmetry (i.e., class AI + $\mathcal{S}_{+}$ in Table~\ref{tab: real AZ + SLS}), each non-Hermitian matrix $h_1$ or $h_2$ respects time-reversal symmetry and belongs to class AI.
According to the analytical result, the distribution of the minimum singular value in $h_1$ or $h_2$ is [see Eq.~(\ref{aeq: s_min_dis_chiral}) in Appendix~\ref{asec: Wigner surmise}]
\begin{align}
    p_{\rm min}^{(h_1)} \left( s \right) 
    = p_{\rm min}^{(h_2)} \left( s \right) 
    = \frac{2+s}{4} e^{-s^2/8 - s/2},
\end{align}
and the cumulative distribution function is
\begin{align}
    F^{(h_i)}_{\rm min} \left( s \right) 
    \coloneqq \int_{s}^{\infty} p_{\rm min}^{(h_i)} \left( t \right) dt 
    = e^{-s\,(4+s)/8}.
\end{align}
Then, the cumulative distribution function $F^{(H)}_{\rm min} \left( s \right)$ for the original non-Hermitian matrix $H$ is obtained as
\begin{align}
    F^{(H)}_{\rm min} \left( s \right) 
    = F^{(h_1)}_{\rm min} \left( s \right) F^{(h_2)}_{\rm min} \left( s \right) 
    = e^{-s\,(4+s)/4},
\end{align}
leading to
\begin{align}
    p_{\rm min}^{(H)} \left( s \right) 
    = - \frac{d}{ds} F^{(H)}_{\rm min} \left( s \right) 
    = \frac{2+s}{2} e^{-s^2/4-s} 
    \neq p_{\rm min}^{(h_i)} \left( s \right).
\end{align}
Still, there is merely a slight difference between $p_{\rm min}^{(H)} \left( s \right)$ and $p_{\rm min}^{(h_i)} \left( s \right)$.
In fact, both of them do not vanish for $s \to 0$ and hence are characterized by $\alpha = 0$.
Additionally, we have $\braket{s_{\rm min}^2}/\braket{s_{\rm min}}^2 = 1.6862 \cdots$ for $p_{\rm min}^{(H)} \left( s \right)$ and $\braket{s_{\rm min}^2}/\braket{s_{\rm min}}^2 = 1.6018 \cdots$ for $p_{\rm min}^{(h_i)} \left( s \right)$, which are nearly equal to each other.
We also numerically obtain $\braket{s_{\rm min}^2}/\braket{s_{\rm min}}^2 = 1.6795 \cdots$ from $5 \times 10^4$ realizations of $10^3 \times 10^3$ non-Hermitian random matrices in the Gaussian ensemble, which is consistent with the analytical result $\braket{s_{\rm min}^2}/\braket{s_{\rm min}}^2 = 1.6862 \cdots$.

\onecolumngrid


\begin{table}[H]
	\centering
	\caption{Singular-value statistics in the complex Altland-Zirnbauer (AZ) symmetry classes of non-Hermitian matrices.
    The complex AZ symmetry classes consist of chiral symmetry (CS) and sublattice symmetry (SLS).
    The subscript of SLS $\mathcal{S}_{\pm}$ specifies the commutation ($+$) or anticommutation ($-$) relation to CS: $\Gamma \mathcal{S}_{\pm} = \pm \mathcal{S}_{\pm} \Gamma$.
    The Hermitized symmetry classes and the corresponding random-matrix indices $\left( \alpha, \beta \right)$ for the singular-value statistics are shown. 
    In the columns of ``$\beta$" and ``$\alpha$", ``N/A (A)" means that the singular-value statistics are not characterized by $\left( \alpha, \beta \right)$ but given by those in Hermitian matrices in class A.}
	\label{tab: complex AZ}
     \begin{tabular}{ccccccc} \hline \hline
    ~~Class~~ & ~~CS~~ & ~~SLS~~ & ~~Classifying space~~ & ~~Hermitization~~ & ~~$\beta$~~ & ~~$\alpha$~~ \\ \hline
    A & $0$ & $0$ & $\mathcal{C}_1$ & AIII & $2$ & $1$ \\
    AIII = A + $\eta$ & $1$ & $0$ & $\mathcal{C}_0$ & A & ~~N/A (A)~~ & ~~N/A (A)~~ \\ \hline
    AIII $+ \mathcal{S}_{+}$ & $1$ & $1$ & $\mathcal{C}_1$ & AIII & $2$ & $1$ \\ \hline
    ~~A $+ \mathcal{S}$ = AIII$^{\dag}$~~ & $0$ & $1$ & $\mathcal{C}_1 \times \mathcal{C}_1$ & ~~AIII $\times$ AIII~~ & N/A (A) & $1$ \\
    AIII $+ \mathcal{S}_{-}$ & $1$ & $1$ & $\mathcal{C}_0 \times \mathcal{C}_0$ & A $\times$ A & N/A (A) & N/A (A) \\ \hline \hline
  \end{tabular}
\end{table}

\begin{table}[H]
	\centering
	\caption{Singular-value statistics in the real Altland-Zirnbauer (AZ) symmetry classes of non-Hermitian matrices.
    The real AZ symmetry classes consist of time-reversal symmetry (TRS), particle-hole symmetry (PHS), and chiral symmetry (CS).
    The Hermitized symmetry classes and the corresponding random-matrix indices $\left( \alpha, \beta \right)$ for the singular-value statistics are shown.
    In the columns of ``$\beta$" and ``$\alpha$", ``N/A (AI)" and ``N/A (AII)" mean that the singular-value statistics are not characterized by $\left( \alpha, \beta \right)$ but given by those in Hermitian matrices in classes AI and AII, respectively.}
	\label{tab: real AZ}
     \begin{tabular}{cccccccc} \hline \hline
    ~~Class~~ & ~~TRS~~ & ~~PHS~~ & ~~CS~~ & ~~Classifying space~~ & ~~Hermitization~~ & ~~$\beta$~~ & ~~$\alpha$~~ \\ \hline
    ~~AI = D$^{\dag}$~~ & +1 & 0 & 0 & $\mathcal{R}_1$ & BDI & $1$ & $0$ \\
    BDI & +1 & +1 & 1 & $\mathcal{R}_2$ & D & $2$ & $0$ \\
    D & 0 & +1 & 0 & $\mathcal{R}_3$ & DIII & $4$ & $1$ \\
    DIII & -1 & +1 & 1 & $\mathcal{R}_4$ & AII & ~~N/A (AII)~~ & ~~N/A (AII)~~ \\
    AII = C$^{\dag}$ & -1 & 0 & 0 & $\mathcal{R}_5$ & CII & $4$ & $3$ \\
    CII & -1 & -1 & 1 & $\mathcal{R}_6$ & C & $2$ & $2$ \\
    C & 0 & -1 & 0 & $\mathcal{R}_7$ & CI & $1$ & $1$ \\
    CI & +1 & -1 & 1 & $\mathcal{R}_0$ & AI & ~~N/A (AI)~~ & ~~N/A (AI)~~ \\ \hline \hline
  \end{tabular}
\end{table}

\begin{table}[H]
	\centering
	\caption{Singular-value statistics in the real Altland-Zirnbauer$^{\dag}$ (AZ$^{\dag}$) symmetry classes of non-Hermitian matrices.
    The real AZ$^{\dag}$ symmetry classes consist of time-reversal symmetry$^{\dag}$ (TRS$^{\dag}$), particle-hole symmetry$^{\dag}$ (PHS$^{\dag}$), and chiral symmetry (CS).
    The Hermitized symmetry classes and the corresponding random-matrix indices $\left( \alpha, \beta \right)$ for the singular-value statistics are shown.
    In the columns of ``$\beta$" and ``$\alpha$", ``N/A (AI)" and ``N/A (AII)" mean that the singular-value statistics are not characterized by $\left( \alpha, \beta \right)$ but given by those in Hermitian matrices in classes AI and AII, respectively.}
	\label{tab: real AZ-dag}
     \begin{tabular}{cccccccc} \hline \hline
    ~~Class~~ & ~~TRS$^{\dag}$~~ & ~~PHS$^{\dag}$~~ & ~~CS~~ & ~~Classifying space~~ & ~~Hermitization~~ & ~~$\beta$~~ & ~~$\alpha$~~ \\ \hline
    AI$^{\dag}$ & +1 & 0 & 0 & $\mathcal{R}_7$ & CI & $1$ & $1$ \\
    BDI$^{\dag}$ & +1 & +1 & 1 & $\mathcal{R}_0$ & AI & ~~N/A (AI)~~ & ~~N/A (AI)~~ \\
    D$^{\dag}$ = AI & 0 & +1 & 0 & $\mathcal{R}_1$ & BDI & $1$ & $0$ \\
    DIII$^{\dag}$ & -1 & +1 & 1 & $\mathcal{R}_2$ & D & $2$ & $0$ \\
    AII$^{\dag}$ & -1 & 0 & 0 & $\mathcal{R}_3$ & DIII & $4$ & $1$ \\
    CII$^{\dag}$ & -1 & -1 & 1 & $\mathcal{R}_4$ & AII & ~~N/A (AII)~~ & ~~N/A (AII)~~ \\
    C$^{\dag}$ = AII & 0 & -1 & 0 & $\mathcal{R}_5$ & CII & $4$ & $3$ \\
    CI$^{\dag}$ & +1 & -1 & 1 & $\mathcal{R}_6$ & C & $2$ & $2$ \\ \hline \hline
  \end{tabular}
\end{table}

\begin{table}[H]
	\centering
	\caption{Singular-value statistics in the real Altland-Zirnbauer (AZ) symmetry classes with sublattice symmetry (SLS).
    The subscript of SLS $\mathcal{S}_{\pm}$ specifies the commutation ($+$) or anticommutation ($-$) relation to time-reversal symmetry (TRS) and/or particle-hole symmetry (PHS).
    For the symmetry classes that involve both TRS and PHS (i.e., classes BDI, DIII, CII, and CI), the first subscript specifies the relation to TRS and the second one to PHS.
    The Hermitized symmetry classes and the corresponding random-matrix indices $\left( \alpha, \beta \right)$ for the singular-value statistics are shown.
    In the columns of ``$\beta$" and ``$\alpha$", ``N/A (A)", ``N/A (AI)", and ``N/A (AII)" mean that the singular-value statistics are not characterized by $\left( \alpha, \beta \right)$ but given by those in Hermitian matrices in classes A, AI, and AII, respectively.}
	\label{tab: real AZ + SLS}
     \begin{tabular}{cccccccc} \hline \hline
    ~~Class~~ & ~~Classifying space~~ & ~~Hermitization~~ & ~~$\beta$~~ & ~~$\alpha$~~ \\ \hline
    BDI + $\mathcal{S}_{++}$ & $\mathcal{R}_1$ & BDI & $1$ & $0$  \\
    DIII + $\mathcal{S}_{--}$ = BDI + $\mathcal{S}_{--}$ & $\mathcal{R}_3$ & DIII & $4$ & $1$ \\
    CII + $\mathcal{S}_{++}$ & $\mathcal{R}_5$ & CII & $4$ & $3$ \\
    CI + $\mathcal{S}_{--}$ = CII + $\mathcal{S}_{--}$ & $\mathcal{R}_7$ & CI & $1$ & $1$ \\ \hline
    AI + $\mathcal{S}_{-}$ = AII + $\mathcal{S}_{-}$ & $\mathcal{C}_1$ & AIII & $2$ & $1$ \\
    BDI + $\mathcal{S}_{-+}$ = DIII + $\mathcal{S}_{-+}$ & $\mathcal{C}_0$ & A & ~~N/A (A)~~ & ~~N/A (A)~~ \\
    D + $\mathcal{S}_{+}$ & $\mathcal{C}_1$ & AIII & $2$ & $1$ \\
    DIII + $\mathcal{S}_{-+}$ = BDI + $\mathcal{S}_{-+}$ & $\mathcal{C}_0$ & A & ~~N/A (A)~~ & ~~N/A (A)~~ \\
    AII + $\mathcal{S}_{-}$ = AI + $\mathcal{S}_{-}$ & $\mathcal{C}_1$ & AIII & $2$ & $1$ \\
    CII + $\mathcal{S}_{-+}$ = CI + $\mathcal{S}_{-+}$ & $\mathcal{C}_0$ & A & ~~N/A (A)~~ & ~~N/A (A)~~ \\
    C + $\mathcal{S}_{+}$ & $\mathcal{C}_1$ & AIII & $2$ & $1$ \\
    CI + $\mathcal{S}_{-+}$ = CII + $\mathcal{S}_{-+}$ & $\mathcal{C}_0$ & A & ~~N/A (A)~~ & ~~N/A (A)~~ \\ \hline
    BDI + $\mathcal{S}_{--}$ = DIII + $\mathcal{S}_{--}$ & $\mathcal{R}_3$ & DIII & $4$ & $1$ \\
    DIII + $\mathcal{S}_{++}$ & $\mathcal{R}_5$ & CII & $4$ & $3$ \\
    CII + $\mathcal{S}_{--}$ = CI + $\mathcal{S}_{--}$ & $\mathcal{R}_7$ & CI & $1$ & $1$ \\
    CI + $\mathcal{S}_{++}$ & $\mathcal{R}_1$ & BDI & $1$ & $0$ \\ \hline
    AI + $\mathcal{S}_{+}$ & $\mathcal{R}_1 \times \mathcal{R}_1$ & ~~BDI $\times$ BDI~~ & ~~N/A (AI)~~ & $0$ \\
    BDI + $\mathcal{S}_{+-}$ & $\mathcal{R}_2 \times \mathcal{R}_2$ & ~~D $\times$ D~~ & ~~N/A (A)~~ & $0$ \\
    D + $\mathcal{S}_{-}$ & $\mathcal{R}_3 \times \mathcal{R}_3$ & ~~DIII $\times$ DIII~~ & ~~N/A (AII)~~ & $1$ \\
    DIII + $\mathcal{S}_{+-}$ & $\mathcal{R}_4 \times \mathcal{R}_4$ & ~~AII $\times$ AII~~ & ~~N/A (AII)~~ & ~~N/A (AII)~~ \\
    AII + $\mathcal{S}_{+}$ & $\mathcal{R}_5 \times \mathcal{R}_5$ & ~~CII $\times$ CII~~ & ~~N/A (AII)~~ & $3$ \\
    CII + $\mathcal{S}_{+-}$ & $\mathcal{R}_6 \times \mathcal{R}_6$ & ~~C $\times$ C~~ & ~~N/A (A)~~ & $2$ \\
    C + $\mathcal{S}_{-}$ & $\mathcal{R}_7 \times \mathcal{R}_7$ & ~~CI $\times$ CI~~ & ~~N/A (AI)~~ & $1$ \\
    CI + $\mathcal{S}_{+-}$ & $\mathcal{R}_0 \times \mathcal{R}_0$ & ~~AI $\times$ AI~~ & ~~N/A (AI)~~ & ~~N/A (AI)~~\\ \hline \hline
  \end{tabular}
\end{table}

\twocolumngrid

\section{Dissipative quantum chaos}
    \label{sec: dissipative quantum chaos}

Eigenvalues of nonintegrable quantum systems isolated from the environment obey the statistics of Hermitian random matrices, characterizing their chaotic behavior~\cite{BGS-84, Huse-review, Rigol-review, Haake-textbook}.
As its generalization to dissipative quantum chaos, we show that singular values of nonintegrable open quantum systems obey the statistics of non-Hermitian random matrices.
To this end, we study the singular-value statistics of nonintegrable Lindbladians (Sec.~\ref{subsec: Lindbladian}) and non-Hermitian Hamiltonians (Sec.~\ref{subsec: non-Hermitian Hamiltonian}).

\begin{figure}[t]
\centering
\includegraphics[width=\linewidth]{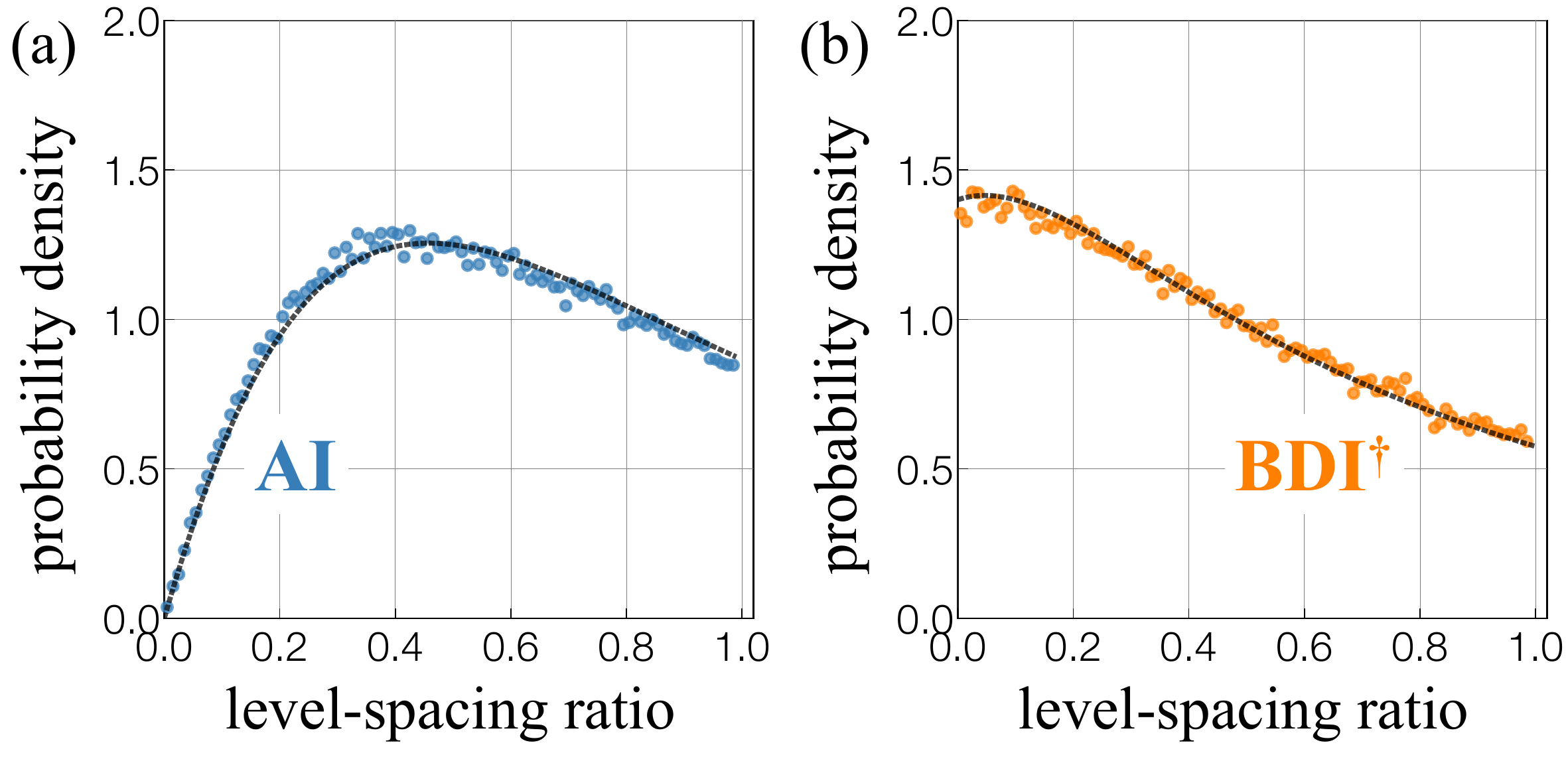} 
\caption{Singular-value statistics of level-spacing ratios for many-body Lindbladians with open boundaries ($L=7$, $J=1.0$, $h_x = -1.05$, $h_z = 0.2$, $\gamma = 0.75$) subject to (a)~damping $L_{n} = \sqrt{\gamma} \sigma_{n}^{-}$ (class AI) and (b)~dephasing $L_{n} = \sqrt{\gamma} \sigma_{n}^{z}$ (class BDI$^{\dag}$).
Singular values of the shifted Lindbladians $\mathcal{L} - \left( \mathrm{tr}\,\mathcal{L}/\mathrm{tr}\,I\right) I$ are considered.
The averages of the singular-value-spacing ratios are (a)~$\braket{r} = 0.529$ and (b)~$\braket{r} = 0.425$.
All the results are taken from singular values away from the spectral edges and averaged over $20$ disorder realizations.
The black dashed curves are the analytical results for small non-Hermitian random matrices in classes (a)~AI [Eq.~(\ref{aeq: Wigner-pr-chiral&BdG}) with $\beta = 1$] and (b)~BDI$^{\dag}$ [Eq.~(\ref{aeq: Wigner-pr-standard}) with $\beta = 1$]}.
	\label{fig: Lindblad}
\end{figure}

\subsection{Lindbladians}
    \label{subsec: Lindbladian}

We investigate open quantum systems described by the quantum master equation $d\rho/dt = \mathcal{L} \left( \rho \right)$ with the Lindbladian~\cite{GKS-76, Lindblad-76, Nielsen-textbook, Breuer-textbook, Rivas-textbook}
\begin{align}
    \mathcal{L} \left( \rho \right) &= -\ii\,[H, \rho] + \sum_{n} \left[ L_n \rho L_{n}^{\dag} - \frac{1}{2}\,\{ L_{n}^{\dag} L_{n}, \rho \} \right],
\end{align}
where $H$ is a Hermitian Hamiltonian for the coherent dynamics, and $L_n$'s are dissipators for the coupling with the surrounding environment.
Here, we choose the Hamiltonian as the quantum Ising model 
\begin{align}
H = - J \sum_{n=1}^{L-1} \left( 1+ \varepsilon_{n} \right) \sigma_{n}^{z} \sigma_{n+1}^{z} - \sum_{n=1}^{L} \left( h_x \sigma_{n}^{x} + h_z \sigma_{n}^{z} \right),
\end{align}
where the open boundary conditions are imposed, and $\varepsilon_{n}$ is randomly chosen from $\left[-0.1, 0.1 \right]$ for each site $n$ to break unwanted symmetry. 
We choose the dissipators as damping 
\begin{align}
L_{n} = \sqrt{\gamma} \sigma_{n}^{-}
\end{align}
or dephasing 
\begin{align}
L_{n} = \sqrt{\gamma} \sigma_{n}^{z}.
\end{align}

We study the singular-value statistics of the Lindbladian $\mathcal{L}$ by doubling the Hilbert space.
Specifically, we map the density operator $\rho = \sum_{i, j} \rho_{ij} \ket{i} \bra{j}$ to a pure state $\ket{\rho} = \sum_{i, j} \rho_{ij} \ket{i} \ket{j}$ in the double Hilbert space.
Through this vectorization procedure $\rho \to \ket{\rho}$, the Lindblad equation reduces to $\left( d/dt \right) \ket{\rho} = \mathcal{L} \ket{\rho}$, where $\mathcal{L}$ is the non-Hermitian operator in the double Hilbert space:
\begin{align}
    &{\cal L} = -\ii \left( {H} \otimes I^{-} - I^{+} \otimes {H}^{*} \right) \nonumber \\
    &~+ \sum_{n} \left[ {L}_{n} \otimes {L}_{n}^{*} - \frac{1}{2} ({L}_{n}^{\dag} {L}_{n} \otimes I^{-}) - \frac{1}{2} (I^{+} \otimes {L}_{n}^{T} {L}_{n}^{*}) \right].
\end{align}
We calculate the spacing ratios of singular values for the Lindbladians $\mathcal{L}$ with the damping and dephasing (Fig.~\ref{fig: Lindblad}).
There, we focus on singular values of the traceless non-Hermitian operator $\mathcal{L} - \left( \mathrm{tr}\,\mathcal{L}/\mathrm{tr}\,I\right) I$ by shifting the Lindbladian $\mathcal{L}$ by a constant $\mathrm{tr}\,\mathcal{L}/\mathrm{tr}\,I$.

Since arbitrary Lindbladians $\mathcal{L}$ are required to preserve Hermiticity of density operators, they are invariant under modular conjugation (see, for example, Ref.~\cite{Kawabata-22}), 
\begin{align}
    \mathcal{J} \mathcal{L} \mathcal{J}^{-1} = \mathcal{L},
        \label{eq: modular conjugation}
\end{align}
where $\mathcal{J}$ is an antiunitary operator that exchanges the bra and ket degrees of freedom,
\begin{align}
    \mathcal{J} \left( O^{+} \otimes O^{-} \right) \mathcal{J}^{-1} = O^{-} \otimes O^{+},\quad \mathcal{J} z \mathcal{J}^{-1} = z^{*},
\end{align}
for arbitrary bosonic operators $O^{\pm}$ and complex numbers $z \in \mathbb{C}$.
Consequently, singular values of the damped Ising model obey the random-matrix statistics in class AI [Fig.~\ref{fig: Lindblad}\,(a)].
As also discussed above, time-reversal symmetry affects complex eigenvalues only around the real axis of the Lindbladian spectrum;
conversely, time-reversal symmetry even changes the local correlations of arbitrary singular values and manifests itself more clearly in the singular-value statistics.

On the other hand, the dephased Ising model additionally respects time-reversal symmetry$^{\dag}$ \begin{align}
    \mathcal{L}^{T} = \mathcal{L}, 
        \label{eq: Lindblad-TRS-dag}
\end{align}
leading to the singular-value statistics in class BDI$^{\dag}$ [Fig.~\ref{fig: Lindblad}\,(b)].
There, chiral symmetry arising from the combination of Eqs.~(\ref{eq: modular conjugation}) and (\ref{eq: Lindblad-TRS-dag}) weakens the level repulsion between singular values even in the chaotic regime.
We emphasize that the Lindbladians $\mathcal{L}$ are represented by sparse matrices because of the locality constraint, as opposed to random matrices.
Nevertheless, the singular-value statistics of $\mathcal{L}$ follow the random-matrix statistics, providing a defining feature of chaos and nonintegrability in open quantum systems.

\subsection{Non-Hermitian Hamiltonians}
    \label{subsec: non-Hermitian Hamiltonian}

\begin{figure*}[t]
\centering
\includegraphics[width=0.8\linewidth]{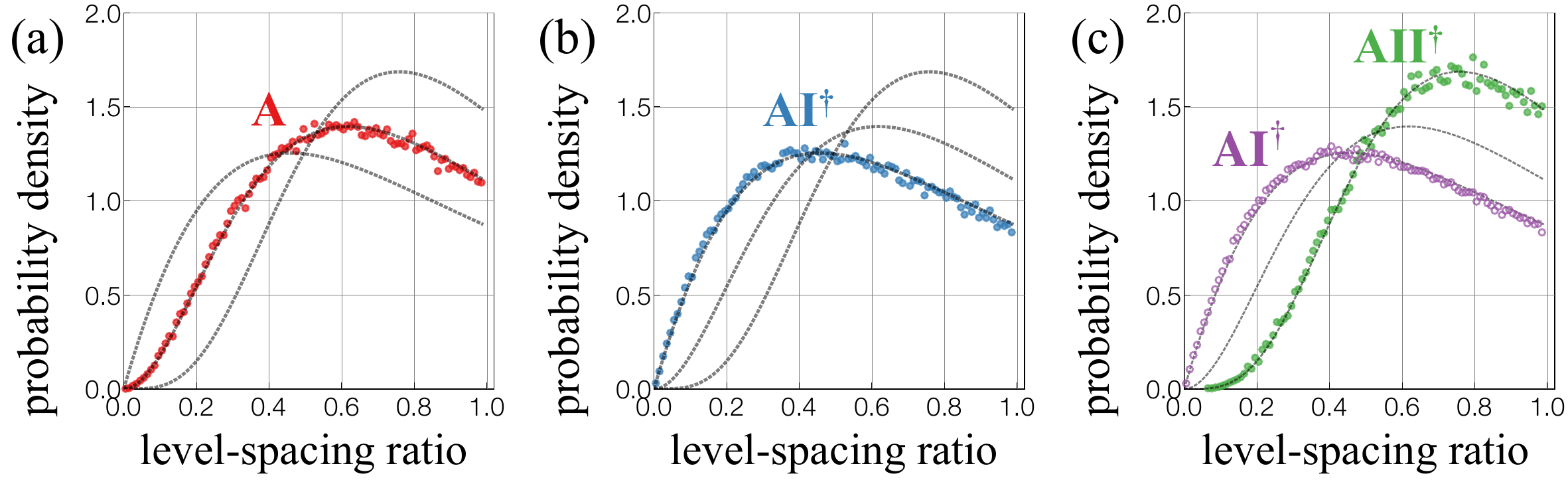} 
\caption{Singular-value statistics of level-spacing ratios for many-body non-Hermitian Hamiltonians under the open boundary conditions ($J=0.2$) for (a)~$h=0.5$, $D=0.9$, $L=13$ (class A; red dots), (b)~$h=0.5$, $D=0$, $L=13$ (class AI$^{\dag}$; blue dots), and (c)~$h=0.0$, $D=0.9$ with $L=13$ (class AII$^{\dag}$; green dots) and $L=14$ (class AI$^{\dag}$; purple dots).
The averages of the level-spacing ratios are (a)~$\braket{r} = 0.600$, (b)~$\braket{r} = 0.531$, and (c)~$\braket{r} = 0.673$ for $L=13$ and $\braket{r} = 0.530$ for $L=14$.
All the results are taken from singular values away from the spectral edges and averaged over $50$ disorder realizations.
The black dashed curves are the
analytical results for small non-Hermitian random matrices
in classes A, AI$^{\dag}$, and AII$^{\dag}$ [Eq.~(\ref{aeq: Wigner-pr-chiral&BdG}) with $\beta = 2, 1, 4$, respectively].}
	\label{fig: NH-spin}
\end{figure*}

In addition to the Lindbladians, we also show that singular values of nonintegrable non-Hermitian Hamiltonians follow the random-matrix statistics in the corresponding symmetry classes.
We study a non-Hermitian spin model~\cite{Hamazaki-19Dyson, Hamazaki-20}
\begin{align}
    H &= - \sum_{n=1}^{L-1} \left( 1 + \ii J \varepsilon_{n} \right) \sigma_{n}^{z} \sigma_{n+1}^{z} - h \sum_{n=1}^{L} \left( -2.1 \sigma_{n}^{x} + \sigma_{n}^{z} \right) \nonumber \\
    &\qquad\qquad\qquad\qquad+ \vec{D} \cdot \sum_{n=1}^{L-1} ( \vec{\sigma}_{n} \times \vec{\sigma}_{n+1} ),
\end{align}
under the open boundary conditions.
Here, $\varepsilon_{n}$ is randomly chosen from $\left[-1, 1 \right]$ for each site $n$ to break unwanted symmetry (e.g., translation symmetry), and $\vec{D} \coloneqq D \left( \vec{e}_x + \vec{e}_z \right)/\sqrt{2}$ is the degree of the Dzyaloshinskii-Moriya interaction. 
Owing to the nonintegrability of this non-Hermitian spin model $H$, singular values obey the random-matrix statistics (Fig.~\ref{fig: NH-spin}).
Depending on the parameters, $H$ belongs to classes A, AI$^{\dag}$, and AII$^{\dag}$, as follows:
\begin{enumerate}
    \item[(i)] For $h \neq 0$ and $D \neq 0$, the non-Hermitian spin model does not respect any symmetries and belongs to class A [Fig.~\ref{fig: NH-spin}\,(a)].

    \item[(ii)] For $h \neq 0$ and $D=0$, the non-Hermitian spin model respects time-reversal symmetry$^{\dag}$
    \begin{align}
        H^{T} = H
    \end{align}
    and belongs to class AI$^{\dag}$ [Fig.~\ref{fig: NH-spin}\,(b)].

    \item[(iii)] For $h=0$ and $D \neq 0$, the non-Hermitian spin model respects time-reversal symmetry$^{\dag}$
    \begin{align}
        \left( \prod_{n=1}^{L} \sigma_{n}^{y} \right) H^{T} \left( \prod_{n=1}^{L} \sigma_{n}^{y} \right)^{-1} = H
    \end{align}
    and belongs to class AI$^{\dag}$ for even $L$ and class AII$^{\dag}$ for odd $L$ [Fig.~\ref{fig: NH-spin}\,(c)].
    For odd $L$, all the singular values are two-fold degenerate.
\end{enumerate}

Time-reversal symmetry$^{\dag}$ also changes the universal level-spacing statistics for complex eigenvalues~\cite{Hamazaki-20}.
Nevertheless, it does not impact the level repulsion for small level spacing but rather influences the peak of the probability distribution.
Conversely, time-reversal symmetry$^{\dag}$ even affects the level repulsion of singular values, as illustrated in Fig.~\ref{fig: NH-spin}.
The distinctive singular-value statistics among the different symmetry classes are clearly discerned according to the power-law behavior for small level spacing, as in Eq.~(\ref{eq: beta}).

\section{Discussions}
    \label{sec: conclusion}

The spectral statistics play a pivotal role in elucidating the nature of chaos and integrability in both closed and open quantum systems, forming the fundamental basis of statistical mechanics.
In this work, we study the statistical properties of singular values, as opposed to complex eigenvalues, in non-Hermitian random matrices, offering a powerful diagnostic tool for characterizing dissipative quantum chaos.
By means of Hermitization, we establish the exhaustive classification of the singular-value statistics for all the 38-fold symmetry classes of non-Hermitian random matrices, summarized as Tables~\ref{tab: complex AZ}-\ref{tab: real AZ + SLS}.
We also analytically obtain the singular-value statistics of small random matrices, 
which well describe those of small random matrices in the similar spirit to the Wigner surmise.
Furthermore, we demonstrate that singular values of nonintegrable Lindbladians and non-Hermitian Hamiltonians conform to the random-matrix statistics, enabling the identification of chaos and nonintegrability in open quantum many-body systems.

Notably, we show that the singular-value statistics provide a useful measure that quantifies dissipative quantum chaos, providing distinct information compared to the complex-eigenvalue statistics.
Specifically, the correlations of complex eigenvalues in the bulk of the spectra are governed solely by time-reversal symmetry$^{\dag}$ in Eq.~(\ref{eq: TRS-dag})~\cite{Hamazaki-20}, while the other symmetries play a role only in the vicinity of the symmetric lines or points~\cite{Xiao-22}, which makes the role of symmetry in dissipative quantum chaos elusive.
By contrast, symmetries irrelevant to the complex-eigenvalue statistics in the bulk, such as time-reversal symmetry in Eq.~(\ref{eq: TRS}), change the spectral statistics of singular values even in the bulk.
Additionally, the difference in the bulk complex-eigenvalue statistics among the symmetry classes is much more subtle compared to Hermitian random matrices~\cite{Hamazaki-20}.
For example, the level repulsion of complex eigenvalues is universally cubic regardless of time-reversal symmetry$^{\dag}$~\cite{Grobe-89}.
Conversely, we show that the level repulsion of singular values relies on symmetry as in Eq.~(\ref{eq: beta}), reminiscent of the Wigner-Dyson universality of Hermitian random matrices.
Thus, the singular-value statistics yield a clearer indicator of symmetry in open quantum systems. 
The combination of the two types of spectral statistics---complex-eigenvalue and singular-value statistics---leads to a more profound understanding of dissipative quantum chaos.

One of the significant applications of the spectral statistics is the identification of phase transitions between the chaotic and integrable regimes, including the Anderson transitions.
In a similar manner to the eigenvalue statistics, the singular-value statistics should be utilized to obtain the universal critical exponents and scaling functions of the Anderson transitions, which we leave for future work. 
In this respect, it is noteworthy that the critical points of the Anderson transitions generally depend on the eigenvalues.
Consequently, to capture the critical behavior precisely, we need to focus on the spectral statistics within a fixed window of eigenvalues.
However, for non-Hermitian systems, the complex-eigenvalue window includes fewer eigenstates since complex eigenvalues are distributed in the two-dimensional complex plane, thereby making the numerical finite-size scaling more challenging~\cite{Luo-21L}.
On the other hand, owing to the real-valued nature of the singular values, even in non-Hermitian systems, qualitatively more singular values are included in a fixed window of singular values.
Thus, the singular-value statistics should be advantageous for precisely characterizing the Anderson transitions in open systems.

Finally, it is also worthwhile to further explore the fundamental role of singular values in the physics of open quantum systems.
The singular-value statistics are also relevant to closed quantum systems.
As a prime example, reflection matrices in the scattering process are generally non-Hermitian, even when the corresponding Hamiltonian is Hermitian.
Accordingly, singular values of reflection matrices, square roots of the reflection probability, describe various quantum transport phenomena, such as mesoscopic electronic transport and nuclear reactions.
Notably, for Hermitian Hamiltonians within the AZ symmetry class, the corresponding non-Hermitian reflection matrices generally belong to the AZ$^{\dag}$ symmetry class, as demonstrated in Appendix~\ref{asec: reflection}.
Thus, the statistical properties of singular values in non-Hermitian reflection matrices are described by our classification tables, especially those for the 10-fold AZ$^{\dag}$ symmetry class in Tables~\ref{tab: complex AZ} and \ref{tab: real AZ-dag}.

\begin{acknowledgments}
K.K. is supported by the Japan Society for the Promotion of Science (JSPS) through the Overseas Research Fellowship, and by the Gordon and Betty Moore Foundation through Grant No.~GBMF8685 toward the Princeton theory program.
Z.X. and R.S. are supported by the National Basic Research Programs of China (No.~2019YFA0308401) and by the National Natural Science Foundation of China (No.~11674011 and No.~12074008).
T.O. is supported by JSPS KAKENHI Grants No.~19H00658 and No.~22H05114, and CREST Grant No.~JPMJCR18T4.
\end{acknowledgments}

\appendix

\begin{table*}[t]
	\centering
	\caption{Singular-value statistics of Hermitian random matrices.
    The 10-fold Altland-Zirnbauer (AZ) symmetry classes consist of time-reversal symmetry (TRS), particle-hole symmetry (PHS), and chiral symmetry (CS).
    For TRS and PHS, the entries ``$\pm 1$" mean the presence of symmetry and its sign, and the entries ``$0$" mean the absence of symmetry. 
    For CS, the entries ``$1$" and ``$0$" mean the presence and absence of symmetry, respectively.
    Each class is characterized by the random-matrix indices $\left( \alpha, \beta \right)$ for the singular-value statistics.
    Both numerical results and analytical results based on the Wigner surmise (WS) 
    (i.e., analytical results for small Hermitian random matrices)
    of the average of level-spacing ratios, $\braket{r}$, and the average normalized variance of the minimum singular value, $\braket{s_{\rm min}^2}/\braket{s_{\rm min}}^2$, are obtained from the Gaussian ensemble.
    All the numerical results are averaged over $10^4$ and $5 \times 10^4$ realizations of $10^3 \times 10^3$ matrices for $\braket{r}$ and $\braket{s_{\rm min}^2}/\braket{s_{\rm min}}^2$, respectively.}
	\label{tab: singular value - Hermitian}
     \begin{tabular}{cccccccccc} \hline \hline
    ~~AZ class~~ & ~~TRS~~ & ~~PHS~~ & ~~CS~~ & ~~$\beta$~~ & ~~$\alpha$~~ & ~~$\braket{r}_{\text{num.}}$~~ & ~~$\braket{r}_{\text{WS}}$~~ & $\left[ \braket{s_{\rm min}^2}/\braket{s_{\rm min}}^2 \right]_{\text{num.}}$ & $\left[ \braket{s_{\rm min}^2}/\braket{s_{\rm min}}^2 \right]_{\text{WS}}$ \\ \hline
    A & $0$ & $0$ & $0$ & ~~N/A~~ & ~~N/A~~ & ~~$0.422245 \cdots$~~ & ~~$0.420601 \cdots$~~ & $1.51134 \cdots$ & $1.4786\cdots$ \\
    AIII & $0$ & $0$ & $1$ & $2$ & $1$ & ~~$0.599681 \cdots$~~ & ~~$0.602658 \cdots$~~ & $1.27376 \cdots$ & $1.27324 \cdots$ \\ \hline
    AI & $+1$ & $0$ & $0$ & N/A & N/A & ~~$0.423589 \cdots$~~ & ~~$0.421018 \cdots$~~ & $1.58809 \cdots$ & $1.5708 \cdots$ \\
    BDI & $+1$ & $+1$ & $1$ & $1$ & $0$ & ~~$0.530654 \cdots$~~ & ~~$0.535898 \cdots$~~ & $1.60343 \cdots$ & $1.6018 \cdots$ \\
    D & $0$ & $+1$ & $0$ & $2$ & $0$ & ~~$0.599653 \cdots$~~ & ~~$0.602658 \cdots$~~ & $1.58237 \cdots$ & $1.57954 \cdots$ \\
    DIII & $-1$ & $+1$ & $1$ & $4$ & $1$ & ~~$0.674358 \cdots$~~ & ~~$0.676168 \cdots$~~ & $1.2766 \cdots$ & $1.27324 \cdots$ \\
    AII & $-1$ & $0$ & $0$ & N/A & N/A & ~~$0.411438 \cdots$~~ & ~~$0.409746 \cdots$~~ & $1.44107 \cdots$ & $1.32916 \cdots$ \\
    CII & $-1$ & $-1$ & $1$ & $4$ & $3$ & ~~$0.674414 \cdots$~~ & ~~$0.676168 \cdots$~~ & $1.13004 \cdots$ & $1.12916 \cdots$ \\
    C & $0$ & $-1$ & $0$ & $2$ & $2$ & ~~$0.59974 \cdots$~~ & ~~$0.602658 \cdots$~~ & $1.17321 \cdots$ & $1.17531 \cdots$ \\
    CI & $+1$ & $-1$ & $1$ & $1$ & $1$ & ~~$0.530768 \cdots$~~ & ~~$0.535898 \cdots$~~ & $1.27165 \cdots$ & $1.27324 \cdots$ \\ \hline \hline 
  \end{tabular}
\end{table*}

\begin{figure*}[t]
\centering
\includegraphics[width=0.7\linewidth]{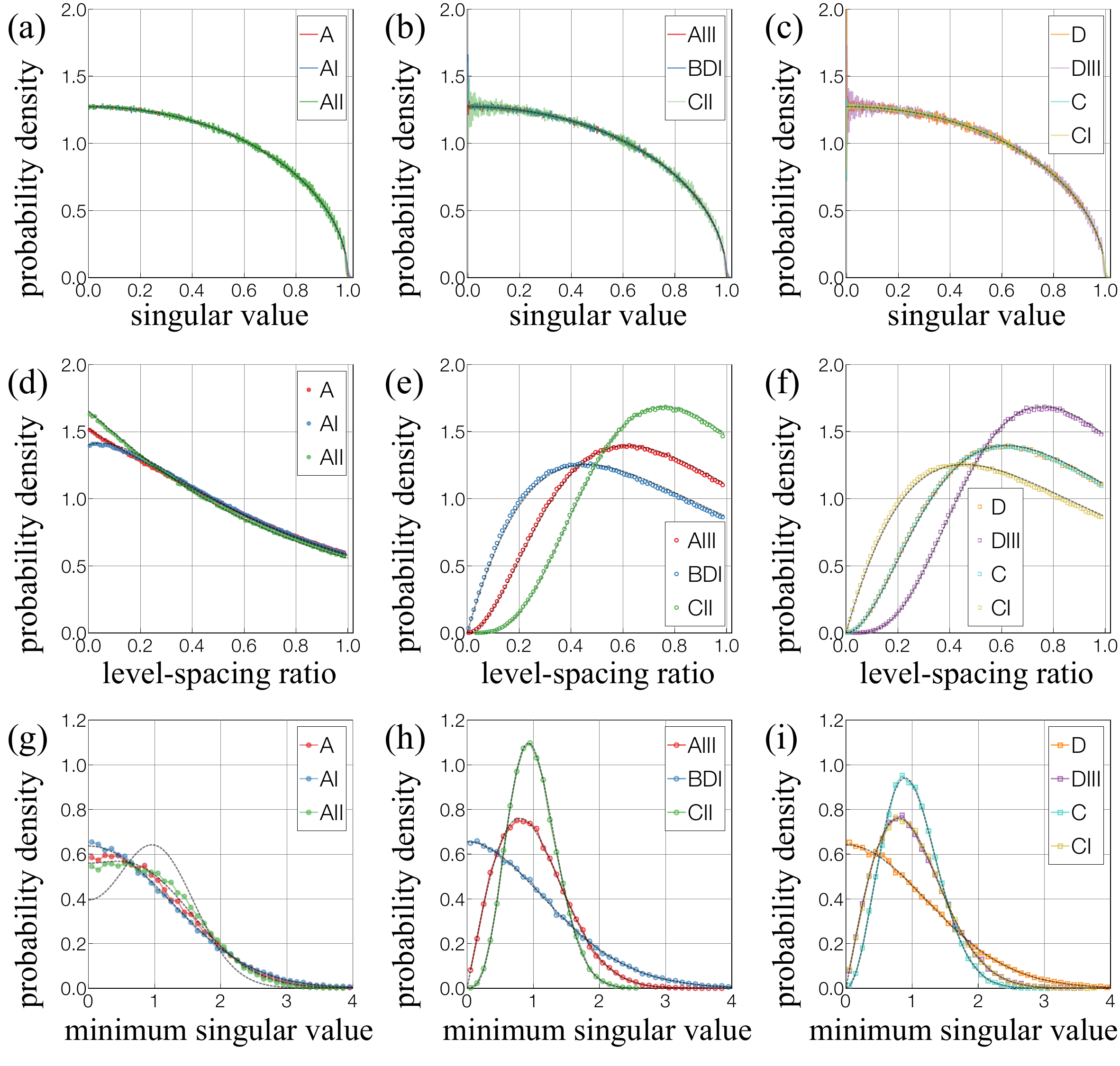} 
\caption{Singular-value statistics of Hermitian random matrices.
All the results are averaged over (a-f)~$10^4$ and (g-i)~$5 \times 10^4$ realizations of $10^3 \times 10^3$ matrices in the Gaussian ensemble. 
(a-c)~Density of singular values $s$ for the (a)~standard classes (classes A, AI, and AII), (b)~chiral classes (classes AIII, BDI, and CII), and (c)~Bogoliubov-de Gennes (BdG) classes (classes D, DIII, C, and CI).
The singular-value spectra are normalized such that their radii are $1$.
The black dashed curves are the Wigner semicircle law $\rho = \left( 4/\pi \right) \sqrt{1-s^2}$.
(d-f)~Level-spacing-ratio distributions for the (d)~standard classes, (e)~chiral classes, and (f)~BdG classes.
The black dashed curves are the analytical results
for small Hermitian random matrices.
(g-i)~Distributions of the minimum singular value $s_{\rm min}$ for the (g)~standard classes, (h)~chiral classes, and (i)~BdG classes.
The probability distribution functions are normalized such that their averages $\braket{s_{\rm min}}$ are $1$.
The black dashed curves are the analytical results
for small Hermitian random matrices.
}
	\label{fig: Hermitian RMT}
\end{figure*}

\section{Singular-value statistics of Hermitian random matrices}
    \label{asec: Hermitian}

As discussed in Sec.~\ref{subsec: Hermitization}, the singular-value statistics of non-Hermitian random matrices reduce to those of Hermitian random matrices.
Here, we study the singular-value statistics of Hermitian random matrices in all the 10-fold AZ symmetry classes (Table~\ref{tab: singular value - Hermitian} and Fig.~\ref{fig: Hermitian RMT}).
Figure~\ref{fig: Hermitian RMT}\,(a-c) shows the density of singular values
whose spectral radius is normalized to be unity.
Away from zero singular value $s=0$, it is described by the Wigner semicircle law 
\begin{align}
    \rho \left( s \right) = \frac{4}{\pi} \sqrt{1-s^2} 
    \label{aeq: semicircle}
\end{align}
for all the symmetry classes.
Equivalently, squared singular values $s^2$ follow the Marchenko-Pastur distribution~\cite{Marchenko-Pastur-67}
\begin{align}
    \rho \left( s^2 \right) = \frac{2}{\pi} \sqrt{\frac{1-s^2}{s^2}}
\end{align}
for square matrices.
Around $s=0$, on the other hand, the density of singular values exhibits different behaviors depending on symmetry classes, as discussed below.

Figure~\ref{fig: Hermitian RMT}\,(d-f) shows the distributions of the level-spacing ratios $r$ of singular values in Eq.~(\ref{eq: spacing ratio}).
The average of $r$ gives a useful measure that distinguishes between different symmetry classes, as summarized in Table~\ref{tab: singular value - Hermitian}.
The singular-value spectra exhibit the two-fold (four-fold) degeneracy in classes AIII, BDI, D, AII, C, and CI (DIII and CII).
In the chiral and BdG classes, eigenvalues come in opposite-sign pairs because of chiral or particle-hole symmetry, and hence the level-spacing-ratio (or level-spacing) distributions $p_{\rm r} = p_{\rm r} \left( r \right)$ of singular values reduce to those of eigenvalues~\cite{Oganesyan-Huse-07, Atas-13}, which are determined only by time-reversal symmetry and described by the Dyson index $\beta$.
In the standard classes, by contrast, $p_{\rm r}$'s are not described by any of the conventional eigenvalue statistics.
For example, the probability density for zero level-spacing ratio, $p_{\rm r} \left( r=0 \right)$, is nonvanishing in the standard classes in contrast to the vanishing probability in the chiral and BdG classes.
This behavior originates from the fact that uncorrelated levels can be close to each other in the singular-value spectrum without chiral or particle-hole symmetry.
In fact, $p_{\rm r}$'s in the standard classes reduce to those of eigenvalues for two independent random matrices.
Notably, a part of the spectral correlations still survives, which leads to the 3-fold distributions with $p_{\rm r} \left( r =0 \right) \neq 0$ but $\braket{r} > \braket{r}_{\rm Poisson} = 2 \log 2 - 1 = 0.38629 \cdots$ for the Poisson distribution in Eq.~(\ref{eq: pr-Poisson}).
In Appendix~\ref{asec: Wigner surmise}, we analytically derive the level-spacing-ratio distributions of singular values for small random matrices.
These analytical results well describe the level-spacing-ratio distributions for large random matrices in the similar spirit to the Wigner surmise.

Figure~\ref{fig: Hermitian RMT}\,(g-i) shows the distributions of the minimum singular value $s_{\rm min}$ 
whose average is normalized to be unity (i.e., $\braket{s_{\rm min}} = 1$).
Here, the average normalized variance of the minimum singular value, $\braket{s_{\rm min}^2}/\braket{s_{\rm min}}^2$, gives a useful measure that distinguishes between different symmetry classes and does not depend on the normalization of singular values.
In the chiral and BdG classes, the distributions of the minimum singular value, $p_{\rm min} = p_{\rm min} \left( s_{\rm min} \right)$, reduce to the distributions of the minimum nonnegative eigenvalue.
In the standard classes, by contrast, $p_{\rm min}$'s do not reduce to any of the eigenvalue statistics. 
In Appendix~\ref{asec: Wigner surmise}, we analytically derive $p_{\rm min}$ for small random matrices in the standard classes.
Importantly, the combination of the level-spacing-ratio (or level-spacing) distributions $p_{\rm r}$ and the distributions $p_{\rm min}$ of the minimum singular value 
completely distinguishes
the 10-fold symmetry classes, each of which is described by the Dyson index $\beta$ in Eq.~(\ref{eq: beta}) and the other random-matrix index $\alpha$ in Eq.~(\ref{eq: alpha}), respectively.
While some symmetry classes can exhibit the same level-spacing-ratio distribution $p_{\rm r}$ (e.g., classes AIII, D, and C), the distributions $p_{\rm min}$ of the minimum singular value are different between different symmetry classes.

\onecolumngrid
\section{Wigner surmise of singular-value statistics}
    \label{asec: Wigner surmise}

\begin{figure*}[t]
\centering
\includegraphics[width=0.7\linewidth]{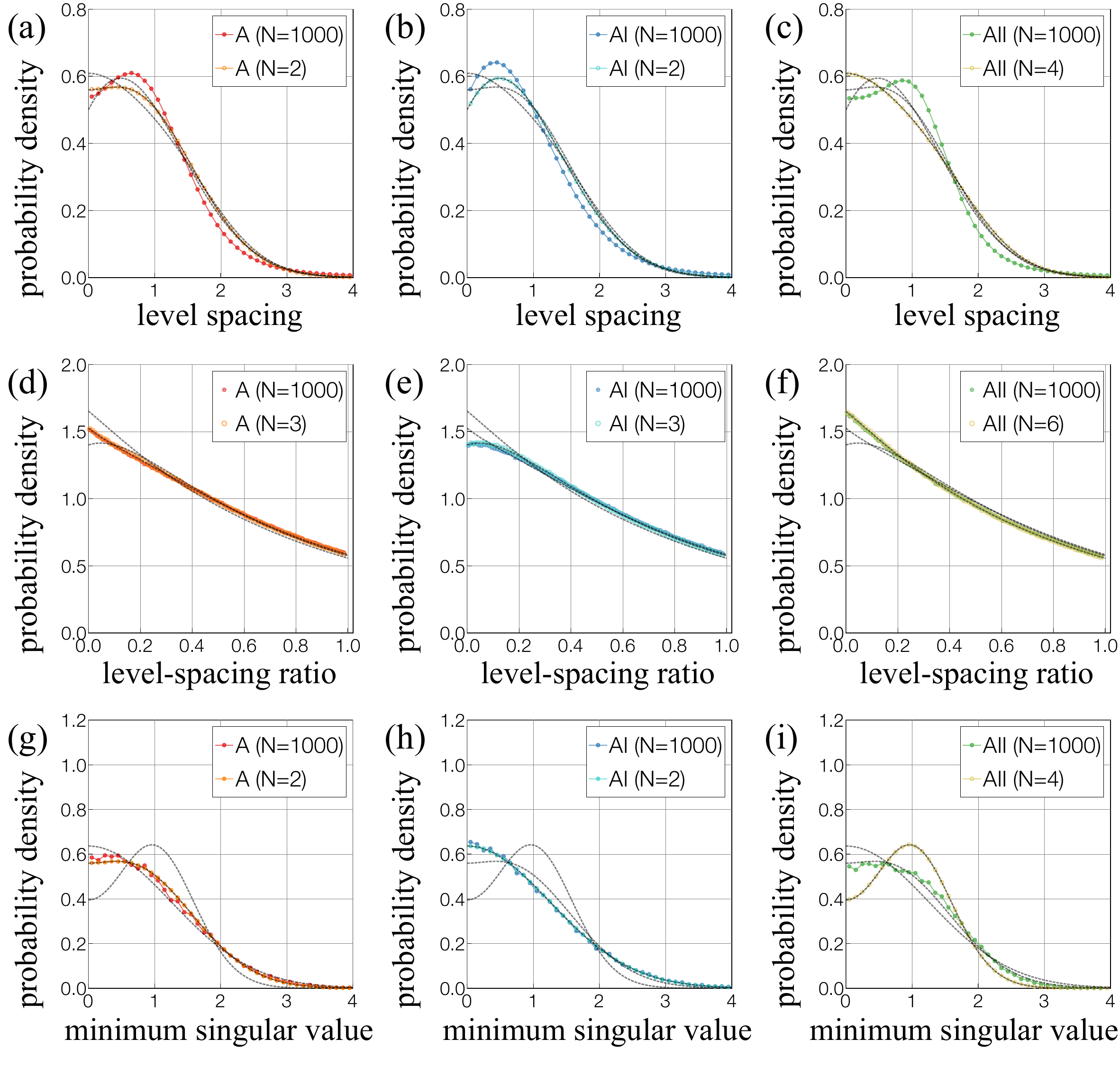} 
\caption{Singular-value statistics of small Hermitian random matrices in the standard classes (i.e., classes A, AI, and AII).
All the numerical results are taken from the Gaussian ensemble and averaged over $10^7$ realizations of $2\times 2$, $3 \times 3$, $4 \times 4$, or $6 \times 6$ matrices and (a-f)~$10^4$ and (g-i)~$5 \times 10^4$ realizations of $10^3 \times 10^3$ matrices.
The black dashed curves are the analytical results.
(a-c)~Level-spacing distributions for classes (a)~A, (b)~AI, and (c)~AII.
(d-f)~Level-spacing-ratio distributions for classes (d)~A, (e)~AI, and (f)~AII.
(g-i)~Distributions of the minimum singular value for classes (g)~A, (h)~AI, and (i)~AII.
The probability distribution functions of the level spacing and minimum singular value are normalized such that their averages are $1$.}
	\label{fig: Wigner surmise}
\end{figure*}

We analytically derive the singular-value statistics of small Hermitian random matrices in all the 10 symmetry classes.
Wigner proposed using small Hermitian random matrices to capture the level-spacing distribution of sufficiently complex nuclei of heavy atoms~\cite{Wigner-51, *Wigner-58}.
Similarly, we show that the analytical results of the singular-value statistics for small random matrices well capture those for large random matrices and also physical systems in the chaotic regime.
As also discussed in Appendix~\ref{asec: Hermitian}, the singular-value statistics in the chiral and BdG classes reduce to the eigenvalue statistics.
In the standard (Wigner-Dyson) classes, by contrast, the singular-value statistics do not reduce to any of the conventional eigenvalue statistics and exhibit the weaker level repulsion than those in the chiral and BdG classes (Fig.~\ref{fig: Wigner surmise}).
In Table~\ref{tab: singular value - Hermitian}, we summarize the average level-spacing ratios $\braket{r}$ and the average normalized variances $\braket{s_{\rm min}^2}/\braket{s_{\rm min}}^2$ of the minimum singular value
for small random matrices,
which well describe the singular-value statistics of large random matrices and serve as useful measures that distinguish between different symmetry classes.

\subsection{Level-spacing distribution}

In the chiral and BdG classes, the level-spacing distributions of singular values reduce to the level-spacing distributions of nonnegative eigenvalues, leading to~\cite{Mehta-textbook, Haake-textbook}
\begin{align}
    p_{\rm s} \left( s \right) = \begin{cases}
       \left( \pi/2 \right) s e^{-\pi s^2/4} & \left( \beta = 1: \text{classes BDI and CI} \right); \\
       \left( 32/\pi^2 \right) s^2 e^{-4 s^2/\pi} & \left( \beta = 2: \text{classes AIII, D, and C} \right); \\
       \left( 262144/729\pi^3 \right) s^4 e^{-64s^2/9\pi} & \left( \beta = 4: \text{classes DIII and CII} \right), \\
    \end{cases}
\end{align}
where $p_{\rm s} \left( s \right)$ is normalized by $\braket{s} = 1$.
We have
\begin{align}
    \frac{\braket{s^2}}{\braket{s}^2} = \begin{cases}
       4/\pi = 1.27324 \cdots & \left( \beta = 1: \text{classes BDI and CI} \right); \\
       3\pi/8= 1.1781 \cdots & \left( \beta = 2: \text{classes AIII, D, and C} \right); \\
       45\pi/128 = 1.10447 \cdots & \left( \beta = 4: \text{classes DIII and CII} \right). \\
    \end{cases}
\end{align}

In the standard classes, the smallest random matrices to study the level-spacing distributions of singular values are $2\times 2$ ($4\times 4$) Hermitian random matrices for classes A and AI (class AII).
The joint probability distribution function of the two eigenvalues in the Gaussian ensemble is given as 
\begin{align}
    \rho \left( \lambda_1, \lambda_2 \right) = N_{\beta} \left| \lambda_1 - \lambda_2 \right|^{\beta} e^{-\lambda_1^2 - \lambda_2^2},
        \label{seq: jpdf-2}
\end{align}
where we assume $\left| \lambda_1 \right| \leq \left| \lambda_2 \right|$, and the normalization constant is 
\begin{align}
    N_{\beta} = \begin{cases}
        \sqrt{2/\pi} & \left( \beta = 1: \text{class AI} \right); \\
        2/\pi& \left( \beta = 2: \text{class A} \right); \\
        2/3\pi & \left( \beta = 4: \text{class AII} \right).
    \end{cases}
\end{align}
The distributions of the level spacing between the two singular values, $\left| \lambda_2 \right| - \left| \lambda_1\right|$, are obtained as
\begin{align}
    p_{\rm s} \left( s \right) 
    &= 2 \int_{0}^{\infty} d\lambda_1  \int_{\lambda_1}^{\infty} d\lambda_2~\rho \left( \lambda_1, \lambda_2 \right) \delta \left( s - \left( \lambda_2 - \lambda_1\right) \right) + 2 \int_{-\infty}^{0} d\lambda_1 \int_{-\lambda_1}^{\infty} d\lambda_2~\rho \left( \lambda_1, \lambda_2 \right) \delta \left( s - \left( \lambda_2 + \lambda_1\right) \right) \nonumber \\
    &= 2 \int_0^{\infty} d\lambda_1~\left[ \rho \left( \lambda_1, \lambda_1 + s \right) + \rho \left( -\lambda_1, \lambda_1 + s \right) \right] \nonumber \\
    &= \begin{cases}
       \sqrt{\cfrac{2}{\pi}} \left( 1 + \sqrt{\cfrac{\pi}{2}} s e^{s^2/2} \mathrm{erfc} \left( s/\sqrt{2}\right) \right) e^{-s^2} & \left( \beta = 1: \text{class AI} \right); \\
        \sqrt{\cfrac{2}{\pi}} \left( \sqrt{\cfrac{2}{\pi}} s + \left( 1+s^2 \right) e^{s^2/2} \mathrm{erfc} \left( s/\sqrt{2}\right) \right) e^{-s^2} & \left( \beta = 2: \text{class A} \right); \\
        \sqrt{\cfrac{2}{\pi}} \left( \cfrac{6s +2s^3}{3\sqrt{2\pi}} + \left( 1 + \cfrac{s^4}{3} \right) e^{s^2/2} \mathrm{erfc} \left( s/\sqrt{2}\right) \right) e^{-s^2} & \left( \beta = 4: \text{class AII} \right).
        \end{cases}
\end{align}
We normalize $p_{\rm s} \left( s \right)$ by $\tilde{p}_{\rm s} \left( s \right) \coloneqq c_{\beta} p_{\rm s} \left( c_{\beta} s \right)$ with
\begin{align}
    c_{\beta} \coloneqq \int_{0}^{\infty} ds~s p_{\rm s} \left( s \right) = \begin{cases}
        \left( 1/2\right) \sqrt{\pi/2} & \left( \beta = 1: \text{class AI} \right); \\
        3 \left( \sqrt{2} - 1 \right)/\sqrt{\pi} & \left( \beta = 2: \text{class A} \right); \\
        \left( 22\sqrt{2} - 23 \right)/6\sqrt{\pi} & \left( \beta = 4: \text{class AII} \right),
    \end{cases}
\end{align}
which satisfies $\int_{0}^{\infty} ds~s \tilde{p}_{\rm s} \left( s \right) = 1$.
We also have
\begin{align}
    \frac{\braket{s^2}}{\braket{s}^2} = \begin{cases}
        8 \left( 2 - \sqrt{2} \right)/\pi = 1.49169 \cdots & \left( \beta = 1: \text{class AI} \right); \\
        2 \left( \pi - 2\right)/9 \left( \sqrt{2} - 1 \right)^2 = 1.4786 \cdots & \left( \beta = 2: \text{class A} \right); \\
        12 \left( 9 \pi - 20 \right)/\left( 22 \sqrt{2} - 23\right)^2 = 1.50863 \cdots & \left( \beta = 4: \text{class AII} \right).
    \end{cases}
\end{align}
Notably, the probability distribution functions do not vanish even for small level spacing $0 \leq s \ll 1$ in contrast with the chiral and BdG classes.
This is because the two singular values can be close to each other in the absence of chiral and particle-hole symmetry even if the corresponding two eigenvalues are away from each other.
Still, a part of the level repulsion survives, which leads to $p_{\rm s} \left( 0 \right) < \left[ p_{\rm s} \left( 0 \right) \right]_{\rm Poisson} = 1$ and $\braket{s^2}/\braket{s}^2 < [ \braket{s^2}/\braket{s}^2 ]_{\rm Poisson} = 2$ with the Poisson statistics $\left[ p_{\rm s} \left( s \right) \right]_{\rm Poisson} = e^{-s}$. 

\subsection{Level-spacing-ratio distribution}

In the chiral and BdG classes, the level-spacing-ratio distributions of singular values reduce to the level-spacing-ratio distributions of nonnegative eigenvalues in the chiral and BdG classes, similarly to the level-spacing distributions, 
leading to~\cite{Atas-13}
\begin{align}
p_{\rm r} \left( r \right) = \frac{1}{Z_\beta} \frac{\left( r+r^2 \right)^{\beta}}{\left( 1+r+r^2 \right)^{1+(3/2)\beta}} \theta \left( 1-r \right)
    \label{aeq: Wigner-pr-chiral&BdG}
\end{align}
with the normalization constants
\begin{align}
    Z_{\beta} = \begin{cases}
       4/27 & \left( \beta = 1: \text{classes BDI and CI} \right); \\
       2\pi/81\sqrt{3} & \left( \beta = 2: \text{classes AIII, D, and C} \right); \\
       2\pi/729\sqrt{3} & \left( \beta = 4: \text{classes DIII and CII} \right).
    \end{cases}
\end{align}
The average level-spacing ratio is obtained as 
\begin{align}
    \braket{r} = \begin{cases}
       4 - 2\sqrt{3} = 0.535898 \cdots & \left( \beta = 1: \text{classes BDI and CI} \right); \\
       2\sqrt{3}/\pi - 1/2 = 0.602658 \cdots & \left( \beta = 2: \text{classes AIII, D, and C} \right); \\
       32/5\sqrt{3}\pi - 1/2 = 0.676168 \cdots & \left( \beta = 4: \text{classes DIII and CII} \right).
    \end{cases}
\end{align}

In the standard classes, the smallest random matrices to study the level-spacing-ratio distributions of singular values are  $3\times 3$ ($6\times 6$) Hermitian random matrices for classes A and AI (class AII).
The joint probability distribution function of the three eigenvalues in the Gaussian ensemble is
\begin{align}
    \rho \left( \lambda_1, \lambda_2, \lambda_3 \right) = N_{\beta} \left| \lambda_1 - \lambda_2 \right|^{\beta} \left| \lambda_2 - \lambda_3 \right|^{\beta} \left| \lambda_3 - \lambda_1 \right|^{\beta} e^{-\lambda_1^2 - \lambda_2^2 - \lambda_3^2},
        \label{eq: jpdf-3}
\end{align}
where we assume $\left| \lambda_1 \right| \leq \left| \lambda_2 \right| \leq \left| \lambda_3 \right|$ and $\lambda_1 \geq 0$, and the normalization constant is 
\begin{align}
    N_{\beta} = \begin{cases}
        \pi/8\sqrt{2} & \left( \beta = 1: \text{class AI} \right); \\
        \pi^{3/2}/8 & \left( \beta = 2: \text{class A} \right); \\
        45 \pi^{3/2}/8 & \left( \beta = 4: \text{class AII} \right).
    \end{cases}
\end{align}

We have four possible cases: 
(i)~$0 \leq \lambda_1 \leq \lambda_2 \leq \lambda_3$; 
(ii)~$0 \leq \lambda_1 \leq -\lambda_2 \leq \lambda_3$; 
(iii)~$0 \leq \lambda_1 \leq \lambda_2 \leq - \lambda_3$;
(iv)~$0 \leq \lambda_1 \leq -\lambda_2 \leq -\lambda_3$.
For example, for the case (i), we have three singular values $0\leq \lambda_1 \leq \lambda_2 \leq \lambda_3$.
Let us introduce $x \coloneqq \lambda_2 - \lambda_1 \geq 0$ and $y \coloneqq \lambda_3 - \lambda_2 \geq 0$, which satisfy $\lambda_2 = \lambda_1 + x$, $\lambda_3 = \lambda_1 + x + y$, and $\int_{0}^{\infty} d\lambda_1 \int_{\lambda_1}^{\infty} d\lambda_2 \int_{\lambda_2}^{\infty} d\lambda_3 = \int_{0}^{\infty} d\lambda_1 \int_{0}^{\infty} dx \int_{0}^{\infty} dy$.
The level-spacing ratio of the singular values is $r = x/y$ for $x \leq y$ and $r = y/x$ for $x \geq y$.
Then, the level-spacing-ratio distribution of the singular values for the case (i) is obtained as
\begin{align}
    &\int_{0}^{\infty} d\lambda_1 \int_{0}^{\infty} dy \int_{0}^{y} dx~\rho \left( \lambda_1, \lambda_1 + x, \lambda_1 + x + y \right) \delta \left(  r - x/y\right) \nonumber \\
    &\qquad\qquad\qquad+ \int_{0}^{\infty} d\lambda_1 \int_{0}^{\infty} dx \int_{0}^{x} dy~\rho \left( \lambda_1, \lambda_1 + x, \lambda_1 + x + y \right) \delta \left(  r - y/x\right) \nonumber \\
    &\qquad= \int_{0}^{\infty} d\lambda_1 \int_{0}^{\infty} dy~\rho \left( \lambda_1, \lambda_1 + ry, \lambda_1 + \left( 1+r \right) y \right) y  
    + \int_{0}^{\infty} d\lambda_1 \int_{0}^{\infty} dx~\rho \left( \lambda_1, \lambda_1 + x, \lambda_1 + \left( 1+r \right) x \right) x.
\end{align}
The other three cases can be evaluated in a similar manner, and the sum of the four terms is given by 
\begin{align}
    p_{\rm r} \left( r \right) &= \int_{0}^{\infty} d\lambda_1 \int_{0}^{\infty} dy \left[ \rho \left( \lambda_1, \lambda_1 + ry, \lambda_1 + \left( 1+r \right) y \right)  + \rho \left( \lambda_1, - \lambda_1 - ry, \lambda_1 + \left( 1+r \right) y \right) \right. \nonumber \\
    &\qquad\qquad \left. + \rho \left( \lambda_1, \lambda_1 + ry, - \lambda_1 - \left( 1+r \right) y \right)  + \rho \left( \lambda_1, - \lambda_1 - ry, - \lambda_1 -  \left( 1+r \right) y \right) \right] y \nonumber \\
    &\qquad + \int_{0}^{\infty} d\lambda_1 \int_{0}^{\infty} dx \left[ \rho \left( \lambda_1, \lambda_1 + x, \lambda_1 + \left( 1+r \right) x \right) + \rho \left( \lambda_1, -\lambda_1 - x, \lambda_1 + \left( 1+r \right) x \right) \right. \nonumber \\
    &\qquad\qquad \left. + \rho \left( \lambda_1, \lambda_1 + x, - \lambda_1 - \left( 1+r \right) x \right) + \rho \left( \lambda_1, -\lambda_1 - x, - \lambda_1 - \left( 1+r \right) x \right) \right] x.
        \label{aeq: Wigner-pr-standard}
\end{align}
For arbitrary $\beta =1, 2, 4$, this integral can be expressed by elementary but complicated functions. 
We have 
\begin{align}
    p_{\rm r} \left( r=0 \right) = \begin{cases}
        \left( 64\sqrt{3} - 40 - 23 \sqrt{2} \right)/8\pi = 1.52488 \cdots & \left( \beta = 1: \text{class AI} \right); \\
        1/2 + 2\sqrt{2}/\pi = 1.40032 \cdots & \left( \beta = 2: \text{class A} \right); \\
        \left( 136192 \sqrt{3} - 100224 - 32481 \sqrt{2} \right)/17280\pi = 1.65293 \cdots & \left( \beta = 4: \text{class AII} \right),
    \end{cases}
\end{align}
and
\begin{align}
    \braket{r} = \begin{cases}
        0.421018 \cdots & \left( \beta = 1: \text{class AI} \right); \\
        0.420601 \cdots & \left( \beta = 2: \text{class A} \right); \\
        0.409746 \cdots & \left( \beta = 4: \text{class AII} \right).
    \end{cases}
\end{align}
While the level repulsion arises even in the singular-value spectrum for the cases (i), (iii), and (iv), it is weakened for the case (ii), which results in the nonzero probability density for $r = 0$.

\subsection{Distribution of the minimum singular value}

In the presence of chiral or particle-hole symmetry, the distributions of the minimum singular value reduce to the distributions of the minimum nonnegative eigenvalue.
In the chiral classes, they are exactly obtained as~\cite{Edelman-88, Forrester-93, Nishigaki-98, *Damgaard-01}
\begin{align}
    p_{\rm min} \left( s \right) = \begin{cases}
        \left( s/2 \right) e^{-s^2/4} & \left( \alpha = 1: \text{class AIII} \right); \\
        \left( 2+s \right) e^{-s^2/8 - s/2}/4 & \left( \alpha = 0: \text{class BDI} \right); \\
        \left( \pi/2 \right)^{1/2} s^{3/2} e^{-s^2/2} I_{3/2} \left( s \right)  & \left( \alpha = 3: \text{class CII} \right),
    \end{cases}
        \label{aeq: s_min_dis_chiral}
\end{align}
where $I_{n} \left( s \right)$ is the modified Bessel function of the first kind.
We then have
\begin{align}
    \frac{\braket{s_{\rm min}^2}}{\braket{s_{\rm min}}^2} = \begin{cases}
       \cfrac{4}{\pi} = 1.27324 \cdots & \left( \alpha = 1: \text{class AIII} \right); \\
       \cfrac{8 - 4 \sqrt{2e\pi}\,\mathrm{erfc} \left( 1/\sqrt{2} \right)}{2e\pi \left( \mathrm{erfc} \left( 1/\sqrt{2} \right)\right)^2} = 1.6018 \cdots & \left( \alpha = 0: \text{class BDI} \right); \\
       \cfrac{2 \left( 2+ \sqrt{2e\pi}\,\mathrm{erf} \left( 1/\sqrt{2} \right)\right)}{e\pi} = 1.12916 \cdots & \left( \alpha = 3: \text{class CII} \right).
    \end{cases}
    \label{aeq: s_min_square}
\end{align}
In the BdG classes, 
we have~\cite{Sun-20}
\begin{align}
    p_{\rm min} \left( s \right) = \begin{cases}
        \left( 1/\pi \right) \left( 6s - 4s^3 + \sqrt{\pi} \left( 3 - 4s^2 + 4s^4 \right) e^{s^2} \mathrm{erfc} \left( s \right)\right) e^{-2s^2} & \left( \alpha = 0: \text{class D} \right); \\
        4s e^{-2s^2} & \left( \alpha = 1: \text{classes DIII and CI} \right); \\
        \left( 2/3\pi \right) s^2 \left( 30s - 4s^3 + \sqrt{\pi} \left( 15 - 12s^2 + 4s^4 \right) e^{s^2} \mathrm{erfc} \left( s \right)\right) e^{-2s^2} & \left( \alpha = 2: \text{class C} \right),
    \end{cases}
        \label{aeq: s_min_dis_BdG}
\end{align}
and
\begin{align}
    \frac{\braket{s_{\rm min}^2}}{\braket{s_{\rm min}}^2} = \begin{cases}
       2 \left( 3\pi - 8\right)/\left( 7-4\sqrt{2} \right)^2 = 1.57954 \cdots & \left( \alpha = 0: \text{class D} \right); \\
       4/\pi = 1.27324 \cdots & \left( \alpha = 1: \text{classes DIII and CI} \right); \\
       2 \left( 15\pi -32 \right)/75 \left( 2-\sqrt{2} \right)^2 = 1.17531 \cdots & \left( \alpha = 2: \text{class C} \right).
    \end{cases}
\end{align}

In the standard classes, the smallest random matrices to study the distribution of the minimum singular value are $2\times 2$ ($4\times 4$) Hermitian random matrices for classes A and AI (class AII).
The joint probability distribution function of the two eigenvalues in the Gaussian ensemble is given as Eq.~(\ref{seq: jpdf-2}), where we assume $\left| \lambda_1 \right| \leq \left| \lambda_2 \right|$.
The distributions of the minimum singular value $\left| \lambda_1 \right|$ are obtained as
\begin{align}
    p_{\rm min} \left( s \right) 
    &= 2 \int_{0}^{\infty} d\lambda_1  \int_{\lambda_1}^{\infty} d\lambda_2~\rho \left( \lambda_1, \lambda_2 \right) \delta \left( s - \lambda_1 \right) + 2 \int_{-\infty}^{0} d\lambda_1 \int_{-\lambda_1}^{\infty} d\lambda_2~\rho \left( \lambda_1, \lambda_2 \right) \delta \left( s + \lambda_1 \right) \nonumber \\
    &= \begin{cases}
        2 \sqrt{\cfrac{2}{\pi}} e^{-2s^2} & \left( \beta = 1: \text{class AI} \right); \\
        \cfrac{2}{\sqrt{\pi}}\left( \cfrac{2s}{\sqrt{\pi}}  + \left( 1 + 2s^2 \right) e^{s^2} \mathrm{erfc} \left( s \right)\right) e^{-2s^2} & \left( \beta = 2: \text{class A} \right); \\
        \cfrac{1}{\sqrt{\pi}} \left( \cfrac{6s + 28s^3}{3\sqrt{\pi}} + \left( 1 + 4s^2 + \cfrac{4}{3} s^4 \right) e^{s^2} \mathrm{erfc} \left( s \right)\right) e^{-2s^2} & \left( \beta = 4: \text{class AII} \right).
        \end{cases}
            \label{aeq: s_min_dis_standard}
\end{align}
We normalize $p_{\rm min} \left( s \right)$ by $\tilde{p}_{\rm min} \left( s \right) \coloneqq c_{\beta} p_{\rm min} \left( c_{\beta} s \right)$ with
\begin{align}
    c_{\beta} \coloneqq \int_{0}^{\infty} ds~s p_{\rm min} \left( s \right) = \begin{cases}
        1/\sqrt{2\pi} & \left( \beta = 1: \text{class AI} \right); \\
        3 \left( 2 - \sqrt{2} \right)/2\sqrt{\pi} & \left( \beta = 2: \text{class A} \right); \\
        \left( 23 - 11 \sqrt{2} \right)/6\sqrt{\pi} & \left( \beta = 4: \text{class AII} \right),
    \end{cases}
\end{align}
which satisfies $\int_{0}^{\infty} ds~s \tilde{p}_{\rm min} \left( s \right) = 1$.
We also have
\begin{align}
    \frac{\braket{s_{\rm min}^2}}{\braket{s_{\rm min}}^2} = \begin{cases}
       2 \left( 3 + 2\sqrt{2} \right) \left( \pi-2 \right)/9 = 1.4785 \cdots & \left( \beta = 1: \text{class AI} \right); \\
       \pi/2 = 1.5707 \cdots & \left( \beta = 2: \text{class A} \right); \\
       6 \left( 771 + 506 \sqrt{2} \right) \left( 9\pi - 16 \right)/82369 = 1.3291 \cdots & \left( \beta = 4: \text{class AII} \right).
    \end{cases}
\end{align}
As shown in Fig.~\ref{fig: Wigner surmise}, while the small-$N$ results well describe the large-$N$ results for classes A and AI, a significant deviation is observed for class AII.

To explain the deviation of the small-$N$ result from the large-$N$ result in class AII, we further calculate the distribution of the minimum singular value for $2m \times 2m$ Hermitian random matrices in class AII with larger sizes ($m > 2$).  
For a generic number $m$ of eigenvalues, the joint probability density function of eigenvalues $\lambda_1$, $\lambda_2$, $\cdots$, $\lambda_m$ in the Gaussian ensemble is
\begin{equation}
    \rho \left(\lambda_1, \lambda_2, \cdots, \lambda_m \right) = N_{4;m} \prod_{i<j} \left( \lambda_i - \lambda_j \right)^4  e^{ - \sum_{i} \lambda_i^2}
        \label{eq: rho-AII}
\end{equation} 
with the normalization constant $N_{4;m} > 0$.
We examine the probability $F \left( s \right)$ of all singular values being larger than $s$, given as
\begin{equation}
    F \left( s \right) = N_{4;m} \int_{|\lambda_1|>s}  \cdots \int_{|\lambda_m|>s} 
    \rho \left( \lambda_1, \lambda_2, \cdots, \lambda_m \right) d\lambda_1 \cdots d\lambda_m \, . 
        \label{F_s}
\end{equation} 
From $F \left( s \right)$, the distribution $p_{\min} \left( s \right)$ of the minimum singular value is obtained as
\begin{equation}
    p_{\min} \left( s \right) = -\frac{d F \left( s \right)}{ds} \,.
\end{equation}
Expanding the term $\prod_{i<j} \left( \lambda_i - \lambda_j \right)^4$ in Eq.~(\ref{eq: rho-AII}) as a polynomial, we find that each term takes the form,
\begin{equation}
   c_{n_1,n_2,\cdots,n_m} \lambda_1^{n_1} \lambda_2^{n_2} \cdots \lambda_m^{n_m} \, .
\end{equation}
Notably, after performing the integral in Eq.~(\ref{F_s}), only the terms with all $n_1$, $\cdots$, $n_m$ being even contribute, and the other terms vanish.
Meanwhile, since $\lambda_i$'s are only dummy variables, we can rewrite $\lambda_1^{n_1} \lambda_2^{n_2} \cdots \lambda_m^{n_m}$ as $\lambda_1^{n^{\prime}_1}  \lambda_2^{n_2^{\prime}} \cdots  \lambda_m^{n_m^{\prime}}$, where $n_1'$, $n_2'$, $\cdots$, $n_m'$ represent the values $n_1, n_2, \cdots, n_m$ arranged in descending order.
The integral can be decomposed into independent Gaussian integrals, each of which is given as
\begin{equation}
    \int_s^{\infty} \lambda^{2n} e^{ -\lambda^2} d\lambda 
    = \frac{1}{2} \int_{s^2}^{\infty} x^{n-\frac{1}{2}} e^{ -x} dx 
    \eqqcolon \frac{1}{2} \Gamma \left( n + \frac{1}{2},s^2 \right)
\end{equation}
with the incomplete gamma function $\Gamma$.
Thus, in principle, we can calculate $F \left( s \right)$ for arbitrary $m$. 
This calculation can be straightforwardly carried out by 
a computer algebra system.
For example, $p_{\min} \left( s \right)$ with $m=3$ is obtained as
\begin{align} 
     &p_{\min} \left( s \right) = -\frac{e^{-3 s^2}}{180 \pi ^{3/2}} \left(-3 \pi  e^{2 s^2} \left(8 \left(2 s^6+4 s^4+15 s^2-15\right) s^2+165\right) \left( \text{erfc} \left( s \right) \right)^2 \right. \nonumber \\
     &\qquad \left.-4 \sqrt{\pi }  s e^{s^2} \left(64 s^{10}+32 s^8+264 s^6-84 s^4-30 s^2+495\right) \text{erfc} \left( s \right)+4 \left(64 s^8+308 s^4-300 s^2-495\right) s^2\right) \, .
\end{align}

\twocolumngrid

\begin{figure}[b]
\centering
\includegraphics[width=\linewidth]{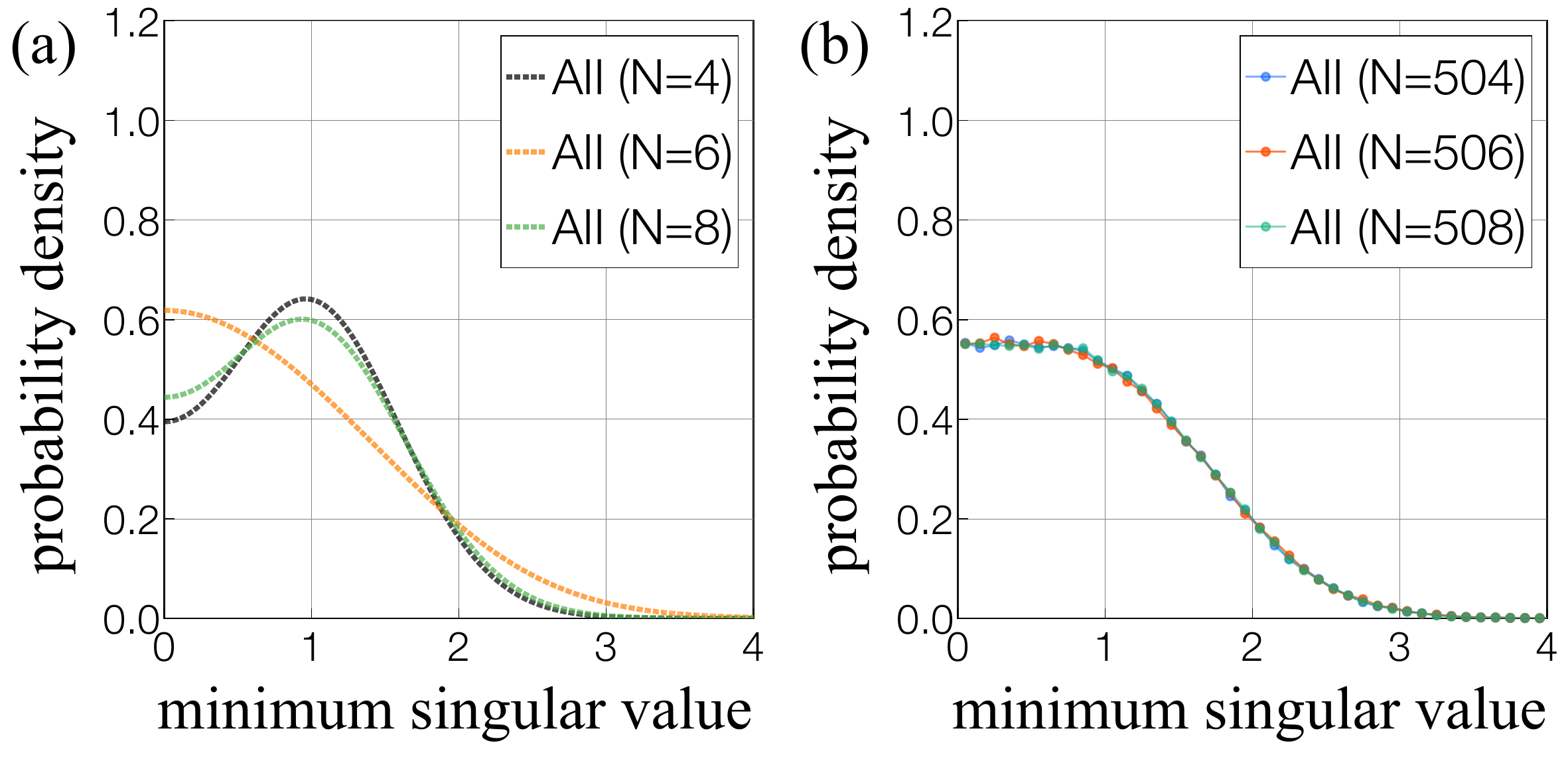} 
\caption{Distributions of the minimum singular value for Hermitian random matrices in class AII.
The probability distribution functions are normalized such that the average of the minimum singular value is $1$.
(a)~Analytical results for $4\times 4$ (black dashed curve), $6\times 6$ (orange dashed curve), and $8\times 8$ (green dashed curve) matrices.
(b)~Numerical results averaged over $2\times 10^5$ realizations of $504\times 504$ (blue dots), $506\times 506$ (red dots), and $508\times 508$ (green dots) matrices.}
	\label{fig: AII}
\end{figure}

\noindent
In this manner, we analytically calculate $p_{\min} \left( s \right)$ for $m=2, 3, 4$ (i.e., $4 \times 4$, $6\times 6$, $8 \times 8$ Hermitian random matrices in class AII) [Fig.~\ref{fig: AII}\,(a)].
Notably, we find a significant even-odd effect depending on the number $m$ of different eigenvalues.
Such an even-odd effect disappears for large random matrices [Fig.~\ref{fig: AII}\,(b)], and $p_{\min} \left( s \right)$ converges to the large-$m$ results in Fig.~\ref{fig: Wigner surmise}\,(i) that lie between the small-even and small-odd results. 

To understand this even-odd effect, let us consider the eigenvalues $\lambda_1$, $\lambda_2$, $\cdots$, $\lambda_{m}$ arranged in ascending order. 
Since the eigenvalue spectrum statistically distributes in a symmetric manner with respect to zero, one can expect that the smallest singular value corresponds to the eigenvalue near the middle of the spectrum. 
Then, for even $m$, the two eigenvalues $\lambda_{m/2}$ and $\lambda_{m/2+1}$ can be the smallest singular value, and the situation for $\lambda_{m/2} < 0$ and $\lambda_{m/2+1} > 0$ is most likely. 
Without loss of generality, we can assume that $\lambda_{m/2+1} > 0$ contributes to the smallest singular value.  
In this scenario, the eigenvalue $\lambda_{m/2+1}$ experiences the asymmetric level repulsion due to the presence of $m/2$ negative eigenvalues and $m/2-1$ positive eigenvalues. 
This asymmetry of the level repulsion becomes significant for small $m$. 
The stronger level repulsion due to the negative eigenvalues prevents the eigenvalue $\lambda_{m/2+1}$ from approaching zero closely, which causes the peak of $p_{\rm min} \left( s \right)$ to shift away from the origin $s=0$ [see Fig.~\ref{fig: AII}\,(a)].
For odd $m$, on the other hand, $\lambda_{\left( m+1 \right)/2}$ lies in the middle of the eigenvalue spectrum and should contribute to the smallest singular value. 
The most likely situation is that half of the remaining $m-1$ eigenvalues are positive while the other half are negative. 
In this scenario, the middle eigenvalue $\lambda_{\left( m+1 \right)/2}$ experiences the equal strength of level repulsion from both the negative and positive eigenvalues. 
Consequently, it tends to approach zero in the eigenvalue spectrum, leading to the peak at the origin in the distribution $p_{\min} \left( s \right)$.
In contrast to class AII, the even-odd effect should not be significant even for small random matrices in classes A and AI owing to the weaker level repulsion.

\section{Normal random matrices}
    \label{asec: normal}

Normal matrices are an important family of matrices. 
A normal matrix $H$ is required to satisfy
\begin{equation}
    [ H, H^{\dagger} ] = 0 
        \label{aeq: normality}
\end{equation}
as its defining property.
Hermitian matrices are a subfamily of normal matrices.
While generic non-Hermitian matrices are not necessarily diagonalized by unitary matrices, a normal matrix $H$ is always diagonalized by a unitary matrix $\U$, 
\begin{equation}
    H = \U \Lambda \U^{\dagger},
\end{equation}
where $\Lambda \coloneqq \mathrm{diag} \left( z_i \right)$ is a diagonal matrix composed of complex eigenvalues $z_i \in \mathbb{C}$. 
One way of generating an ensemble of normal matrices is to start from an ensemble of generic random matrices and selectively include normal matrices while excluding nonnormal ones.  
A numerical algorithm for generating such an ensemble of normal random matrices was provided, for example, in Ref.~\cite{oas1997normal}.
The joint probability distribution function of eigenvalues for normal random matrices without any symmetry (i.e., class A) in the Gaussian ensemble was shown to be identical to that of generic non-Hermitian random matrices~\cite{oas1997normal, chau1998structure, elbau2005density}.
Consequently, generic and normal random matrices share the same eigenvalue-spacing and eigenvalue-spacing-ratio statistics.

In contrast with the complex-eigenvalue statistics,
the singular-value statistics exhibit distinctions between generic and normal random matrices.
While singular values and eigenvalues are not related to each other in generic non-Hermitian matrices, singular values $s_i$'s of a normal matrix are given as the absolute values of its eigenvalues, 
\begin{equation}
    s_i = \left| z_i \right|.
\end{equation}
Then, since the eigenvalues $z_i$'s of a normal matrix are distributed uniformly in a circle in the complex plane,  the density of the singular values $s_i$'s is given as 
\begin{equation}
    \rho \left( s \right) = 2 s \quad \left( 0 \leq s \leq 1 \right),
\end{equation}
with the spectral radius normalized to unity.
Remarkably, this clearly deviates from the semicircle law in Eq.~(\ref{aeq: semicircle}), which is respected for generic non-Hermitian random matrices.

Furthermore, spacings and spacing ratios of singular values follow the Poisson statistics for normal random matrices. 
For two adjacent singular values, $s_i$ and $s_j$, the corresponding eigenvalues, $z_i=s_i e^{i\theta_i}$ and $z_j = s_j e^{i\theta_j}$, can be well separated from each other due to different phases, $\theta_i$ and $\theta_j$. 
Specifically, the probability that $z_i$ and $z_j$ are adjacent complex eigenvalues, which entails $\left| z_i - z_j \right| < \delta$ with $\delta > 0$ being a constant on the order of the mean level spacing of complex eigenvalues, is proportional to $\delta$. 
Given that $\delta$ is of the order $\mathcal{O} \left( N^{-1/2} \right)$ for the matrix size $N$, the probability scales as $\mathcal{O} \left( N^{-1/2} \right)$ and vanishes entirely as $N \rightarrow \infty$, suggesting the vanishing 
repulsion between the adjacent singular values.  
Notably, this situation differs from the Hermitian case, where the repulsion between singular values similarly weakens but a part of it persists. 
This difference arises because Hermiticity makes eigenvalues $z_i$'s real and enforces $\theta_i = 0$ or $\theta_i = \pi$. 
In the presence of such a constraint, the probability of $\left| z_i-z_j \right|< \delta$ scales as $\mathcal{O} \left( 1 \right)$, leading to the deviations from the Poisson statistics.

It is also worthwhile to study the singular-value statistics of normal random matrices in the presence of symmetry.
As an illustrative example, we here study $2 \times 2$ real normal random matrices (i.e., class AI) and show that time-reversal symmetry has a significant impact on the singular-value statistics.   
A generic $2 \times 2$ real matrix $H$ can be parameterized as 
\begin{equation}
    H = a_0 \sigma_0 + a_{x} \sigma_x + \ii a_{y} \sigma_y + a_z \sigma_z
\end{equation}
with $a_{i} \in \mathbb{R}$ and Pauli matrices $\sigma_i$ ($i = 0, x, y, z$).
The normality condition in Eq.~(\ref{aeq: normality}) requires the constraints 
$a_x a_y = 0$ and $a_y a_z = 0$,
further leading to (i)~$a_x = a_z = 0$ and $a_y \neq 0$, or (ii)~$a_y = 0$.
Within the ensemble of normal random matrices induced from the larger ensemble, the probability of the case~(ii) is much higher than that of the case~(i) since fine-tuning of the additional parameter is necessary for the case~(i).
Thus, in the ensemble of normal random matrices with time-reversal symmetry, almost all matrices are required to respect Hermiticity, resulting in the distinct statistics of eigenvalues and singular values.  
We leave more systematic exploration on the interplay of normality and symmetry for future study.

\section{Symmetry classification of non-Hermitian reflection matrices}
    \label{asec: reflection}

The singular-value statistics are also relevant to the physics of closed quantum systems.
As a prime example, reflection matrices in the scattering process are generally non-Hermitian, even when the corresponding Hamiltonian is Hermitian.
Singular values of these reflection matrices, square roots of the reflection probability, describe quantum transport phenomena.
The fluctuations of singular values from sample to sample, or the suppression thereof, are reflected in the conductance fluctuations~\cite{Lee-85, *Lee-87, Altshuler-85, Imry-86, Altshuler-86}.
Here, we show that if original Hermitian Hamiltonians belong to the AZ symmetry class, the corresponding non-Hermitian reflection matrices generally belong to the AZ$^{\dag}$ symmetry class.
Specifically, we establish that time-reversal symmetry in Eq.~(\ref{eq: TRS}) and particle-hole symmetry in Eq.~(\ref{eq: PHS}) of Hamiltonians impose time-reversal symmetry$^{\dag}$ in Eq.~(\ref{eq: TRS-dag}) and particle-hole symmetry$^{\dag}$ in Eq.~(\ref{eq: PHS-dag}) on reflection matrices.
Similarly, chiral symmetry in Eq.~(\ref{eq: SLS}) of Hamiltonians leads to chiral symmetry in Eq.~(\ref{eq: CS}) of reflection matrices.
Thus, the singular-value statistics of non-Hermitian reflection matrices are described by our classification tables for the 10-fold AZ$^{\dag}$ symmetry class (Tables~\ref{tab: complex AZ} and \ref{tab: real AZ-dag}).

The following discussions are based on the Mahaux-Weidenm\"uller formula~\cite{Mahaux-Weidenmuller-68, *Mahaux-Weidenmuller-69, Verbaarschot-85, Beenakker-review-97, *Beenakker-review-15},
\begin{equation}
    S \left( E \right) = \frac{1-\ii \pi K \left( E \right)}{1 + \ii \pi K \left( E \right)},\quad K \left( E \right) \coloneqq W^{\dag} \frac{1}{E-H} W,
\end{equation}
where $E \in \mathbb{R}$ is a single-particle energy of the incident and scattered waves, and $W$ is a coupling matrix between the system and the ideal leads that is assumed to share the same symmetries as the system.
The scattering matrix $S$ reads~\cite{Datta-textbook, Imry-textbook},
\begin{align}
    S \coloneqq \begin{pmatrix}
        r_{\rm L} & t_{\rm L} \\
        t_{\rm R} & r_{\rm R}
    \end{pmatrix},
\end{align}
where $r_{\rm L}$ ($r_{\rm R}$) is a reflection matrix from the left to the left (from the right to the right), and $t_{\rm R}$ ($t_{\rm L}$) is a transmission matrix from the left to the right (from the right to the left).
If the Hamiltonian respects Hermiticity ($H^{\dag} = H$), the scattering matrix respects unitarity ($SS^{\dag} = S^{\dag} S = 1$).
We stress that the reflection matrix $r$ is generally non-Hermitian even if the corresponding Hamiltonian $H$ is Hermitian.
In Ref.~\cite{Zirnstein-21L, *Zirnstein-21B}, the Mahaux-Weidenm\"uller formula is used to understand the bulk-boundary correspondence of non-Hermitian Hamiltonians.
Here, we instead consider Hermitian Hamiltonians and study symmetry of the corresponding non-Hermitian reflection matrices.
When Hamiltonians lack Hermiticity, reflection matrices are no longer described solely by the AZ$^{\dag}$ symmetry class but are still categorized into one of the 38 symmetry classes.

\subsection{Time-reversal symmetry}

Suppose that Hermitian Hamiltonians respect time-reversal symmetry in Eq.~(\ref{eq: TRS}), leading to 
\begin{align}
    \mathcal{T} K^{T} \left( E \right) \mathcal{T}^{-1} = K \left( E \right),
\end{align}
and 
\begin{align}
    \mathcal{T} S^{T} \left( E \right) \mathcal{T}^{-1} = S \left( E \right).
        \label{aeq: T-Smat}
\end{align}
While $W$ typically satisfies $\mathcal{T} W^* \mathcal{T}^{-1} = W$ for ideal leads, we only need $\mathcal{T} W^* \mathcal{T}^{-1} = e^{\ii \theta} W$ with arbitrary $\theta \in \mathbb{R}$ to have these relations (this is also the case for particle-hole and chiral transformations below).
Thus, we have 
\begin{align}
    \mathcal{T} r^{T} \left( E \right) \mathcal{T}^{-1} = r \left( E \right),
        \label{aeq: T-rmat}
\end{align}
which means that the reflection matrix $r$ respects time-reversal symmetry$^{\dag}$ in Eq.~(\ref{eq: TRS-dag}) for arbitrary $E \in \mathbb{R}$.
It should be noted that the matrix sizes of $\mathcal{T}$ are different between Eqs.~(\ref{aeq: T-Smat}) and (\ref{aeq: T-rmat}).
While $\mathcal{T}$ in Eq.~(\ref{aeq: T-Smat}) acts on the incoming and outgoing modes on both left and right sides, $\mathcal{T}$ in Eq.~(\ref{aeq: T-rmat}) acts solely on those on either left or right side.

\subsection{Particle-hole symmetry}

Suppose that Hamiltonians respect particle-hole symmetry in Eq.~(\ref{eq: PHS}), leading to
\begin{align}
    \mathcal{C} K^{*} \left( E \right) \mathcal{C}^{-1} = W^{\dag} \frac{1}{E+H} W = - K \left( -E \right),
\end{align}
and 
\begin{align}
    \mathcal{C} S^{*} \left( E \right) \mathcal{C}^{-1} = S \left( -E \right).
\end{align}
Thus, for zero modes $E=0$, we have
\begin{align}
    \mathcal{C}\,  r^{*} \left( E=0 \right) \mathcal{C}^{-1} = r \left( E=0 \right),
\end{align}
which means that the reflection matrix $r$ respects time-reversal symmetry, or equivalently, $\ii r$ respects particle-hole symmetry$^{\dag}$ in Eq.~(\ref{eq: PHS-dag}).
It should be noted that particle-hole symmetry$^{\dag}$ is respected only for zero modes $E=0$ and is explicitly broken for $E\neq 0$.

\subsection{Chiral symmetry}

Suppose that Hamiltonians respect chiral symmetry in Eq.~(\ref{eq: CS}), leading to
\begin{align}
    \mathcal{S} K \left( E \right) \mathcal{S}^{-1} = W^{\dag} \frac{1}{E+H} W = - K \left( -E \right),
\end{align}
and 
\begin{align}
    \mathcal{S} S^{\dag} \left( E \right) \mathcal{S}^{-1} = S \left( -E \right).
\end{align}
Thus, for zero modes $E=0$, we have
\begin{align}
    \mathcal{S}\,  r^{\dag} \left( E=0 \right) \mathcal{S}^{-1} = r \left( E=0 \right),
\end{align}
which means that the reflection matrix $r$ respects pseudo-Hermiticity, or equivalently, $\ii r$ respects chiral symmetry in Eq.~(\ref{eq: CS}).

\bibliography{NH_RMT.bib}

\end{document}